\documentclass[twocolumn,superscriptaddress,amsmath,amssymb,floatfix,aps,prx,longbibliography]{revtex4-1}
\usepackage{graphics,graphicx}
\usepackage{epstopdf}

\usepackage[usenames]{color}

\begin{document}

\title{Statistical mechanics of thin spherical shells}

\author{Andrej Ko\v{s}mrlj}
\email{andrej@princeton.edu}
\affiliation{Department of Mechanical and Aerospace Engineering, Princeton University, Princeton, NJ 08544}

\author{David R. Nelson}
\email{nelson@physics.harvard.edu}
\affiliation{Department of Physics, Department of Molecular and Cellular Biology and School of Engineering and Applied Sciences, Harvard University, Cambridge, Massachusetts 02138}

\date{\today}

\begin{abstract}

We explore how thermal fluctuations affect the mechanics of thin amorphous spherical shells. In flat membranes with a shear modulus, thermal fluctuations increase the bending rigidity and reduce the in-plane elastic moduli in a scale-dependent fashion. This is still true for spherical shells. However, the additional coupling between the shell curvature, the local in-plane stretching modes and the local out-of-plane undulations, leads to novel phenomena. In spherical shells thermal fluctuations  produce a radius-dependent negative effective surface tension, equivalent to applying an inward external pressure. By adapting renormalization group calculations to allow for a spherical background curvature, we show that while small spherical shells are stable, sufficiently large shells are crushed by this thermally generated ``pressure''. Such shells can be stabilized by an outward osmotic pressure, but the effective shell size grows non-linearly with increasing outward pressure, with the same universal power law exponent that characterizes the response of fluctuating flat membranes to a uniform tension.

\end{abstract}
\maketitle

\section{Introduction}

Continuum elastic theories for plates~\cite{love888,foppl907,vonKarman910} and shells~\cite{sanders63,koiter66} have been under development for over a century, but they are still actively explored, because of the ``extreme mechanics'' generated by geometrical nonlinearities~\cite{krieger12, stoop15}. Initially, these theories were applied to the mechanics of thin macroscopic structures, where the relevant elastic constants (a Young's modulus and a bending rigidity) are related to the bulk material properties and the plate or shell thickness. However, these theories have also been successfully applied to describe mechanical properties of microscopic structures, such as viral capsids~\cite{lidmar03,ivanovska04,michel06,klug06}, bacterial cell walls~\cite{yao99,wang10,nelson12,amir14}, membranes of red blood cells~\cite{waugh79,evans83,park10}, and hollow polymer and polyelectrolyte capsules~\cite{gao01,gordon04,lulevich04,elsner06,zoldesi08}. Note that in these more microscopic examples, the effective elastic constants are not related to bulk mechanical properties, but instead depend on details of microscopic molecular interactions.

At the microscopic scale, thermal fluctuations become important and their effects on flat two dimensional solid membranes have been studied extensively, starting in the late 1980's. Unlike long one dimensional polymers, which perform self-avoiding random walks~\cite{gennesBpolymers,doiB}, arbitrarily large two dimensional membranes remain flat at low temperatures due to the strong thermal renormalizations triggered by flexural phonons,~\cite{nelson87} which result in strongly scale-dependent enhanced bending rigidities and reduced in-plane elastic constants.~\cite{nelsonB, katsnelsonB}. A related scaling law for the membrane structure function of a solution of spectrin skeletons of red blood cells was checked in an \emph{ensemble-averaged} sense via elegant X-ray and light scattering experiments.~\cite{schmidt93}  However, recent advances in growing and isolating free-standing layers of crystalline materials such as graphene, boron nitride or transition metal dichalcogenides~\cite{novoselov05} (not adsorbed onto a bulk substrate or stretched across a supporting structure) hold great promise for exploring how flexural modes affect the mechanical properties of \emph{individual} sheet polymers that are atomically thin. Recent experiments with graphene have in fact observed a $\sim$$4000$-fold enhancement of the bending rigidity,~\cite{blees15} and a reduced Young's modulus~\cite{nicholl15}, although these results may also be influenced by quenched random disorder (e.g., ripples or grain boundaries), which can compete with thermal fluctuations to produce similar effects~\cite{radzihovsky91,kosmrlj13,kosmrlj14}.

\begin{figure}[t]
\includegraphics[scale=.5]{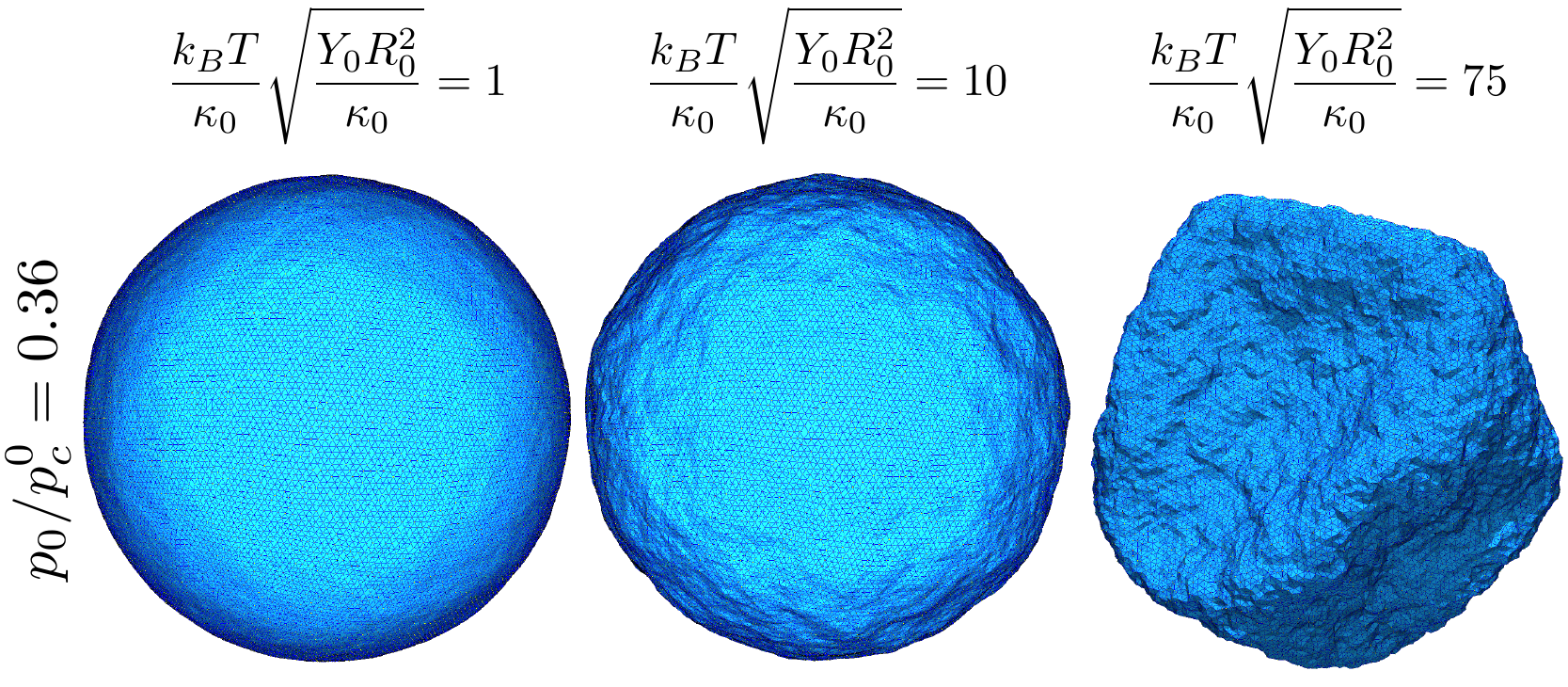}
\caption{(Color online) Snapshots of thermalized spheres from Monte Carlo simulations under inward external pressure $p_0$ at $36\%$ of the classical buckling pressure $p_c^0$ at varying temperatures $T$. All three snapshots are for identical amorphous spherical shells with size $R_0=55 a$ ($a$ is the average mesh size) with bending rigidity $\kappa_0=50 \epsilon$ and Young's modulus $Y_0=577\epsilon/a^2$ ($\epsilon$ sets the energy scale); the F\"oppl-von Karman number characterizing the nonlinear shell mechanics is $\gamma=Y_0 R_0^2/\kappa_0 \approx 35,000$. Shells are under the same inward external pressure $p_0 = 0.08 \epsilon/a^3$, but they are at different temperatures $k_B T=0.267 \epsilon$,  $k_B T=2.67 \epsilon$, $k_B T=20 \epsilon$ (from left to right). Note that the shell on the right is crushed even though the inward external pressure pressure $p_0 \approx 0.36 p_c^0$ is much lower than the classical buckling pressure $p_c^0 = 4 \sqrt{\kappa_0 Y_0}/R_0^2$.
Images are courtesy of Gerrit Vliegenthart and details of simulations are provided in Ref.~\cite{paulose12}.
}
\label{fig:simulations}
\end{figure}

While thermal fluctuations of flat solid sheets are well understood, many microscopic membranes correspond to closed \emph{shells}, and much less is known about their response to thermal fluctuations. The simplest possible shell is an amorphous spherical shell. This was studied by Paulose \emph{et al.}~\cite{paulose12}, where \emph{perturbative} corrections to elastic constants at low temperatures and external pressures were derived and tested with Monte Carlo simulations. Remarkably, these simulations found that at high temperatures thermalized spheres begin to collapse at less than half the classical buckling pressure  (see Fig.~\ref{fig:simulations}). However, it was not possible to quantify this effect, because the perturbative corrections diverge with shell radius. Here, we go well beyond perturbation theory by employing renormalization group techniques, which enable us to study spherical shells over a wide range of sizes, temperatures and external pressures. We show that while spherical shells retain some features of flat solid sheets, there are remarkable new phenomena, such as a thermally generated negative tension, which spontaneously crushes large shells even in the absence of external pressure. We find that shells can be crushed by thermal fluctuations even in the presence of a stabilizing outward pressure!

In Sec.~\ref{sec:elastic_energy}, we review the shallow-shell theory description of thin elastic spheres,~\cite{sanders63,koiter66} while in Sec.~\ref{sec:thermal_fluctuations} we show how to set up the statistical mechanics leading to the thermal shrinkage and fluctuations in the local displacement normal to the shell. Low temperature, perturbative corrections to quantities such as the effective pressure $p$ (a sum of conventional and osmotic contributions), bending rigidity $\kappa$ and Young's modulus $Y$ diverge like $\sqrt{\gamma}$, where $\gamma = Y_0 R_0^2/\kappa_0$ is the F\"oppl-von Karman number of the shell with radius $R_0$ and microscopic elastic moduli $Y_0$ and $\kappa_0$.~\cite{paulose12} A momentum shell renormalization group is then implemented directly for shells embedded in $d=3$ dimensions to resolve these difficulties in Sec.~\ref{sec:renormalization_group}. 
At small scales the bending rigidity and Young's modulus renormalize like flat sheets; however, at large scales the curvature of the shell produces significant changes.  At low temperatures ($k_B T \sqrt{\gamma}/\kappa_0 \ll 1$) the renormalization is cut off already at the elastic length $\ell_\textrm{el}=\left(\kappa_0 R_0^2/Y_0\right)^{1/4}$. At large temperatures  ($k_B T \sqrt{\gamma}/\kappa_0 \gg 1$) and beyond an important thermal length scale $\ell_\textrm{th}\sim \kappa_0/\sqrt{k_B T Y_0}$, the bending rigidity and Young's modulus renormalize with length scale $\ell$ like flat sheets with $\kappa_R \approx \kappa_0 (\ell/\ell_\textrm{th})^\eta$ and $Y_R \approx Y_0 (\ell_\textrm{th}/\ell)^{\eta_u}$, where $\eta \approx 0.8$ and $\eta_u \approx 0.4$.~\cite{aronovitz88} However, this renormalization is interrupted as one scales out to the shell radius $R_0$. For zero pressure we find that shells become unstable to a finite wave-vector mode appearing at the scale $\ell^* \sim \ell_\textrm{th} [\ell_\textrm{el}/\ell_\textrm{th}]^{4/(2+\eta)} \propto R_0^{2/(2+\eta)}\ll R_0$. A sufficiently large  (negative) outward pressure stabilizes the shell and leads to an alternative infrared cut off given by a pressure-dependent length scale $\ell_p$. Detailed results for correlation functions, renormalized couplings and the change in the shell radius can be obtained by integrating the renormalization group flow equations out to scales where the thermal averages are no longer singular. In Sec.~\ref{sec:renormalization_group}, we also present a simple, intuitive derivation of the scaling relation $\eta_u + 2 \eta=2$, originally derived using Ward identities associated with rotational invariance in Ref.~\cite{aronovitz88,guitter89}. In Sec.~\ref{sec:buckling_pressure}, we use the renormalization group method to study the dependence of the renormalized buckling pressure $p_{c}$ on temperature, shell radius and the elastic parameters, which defines a limit of metastability for thermalized shells. The calculated scaling function $\Psi(x)$ defined by $p_{c} = p_c^0 \Psi(k_B T \sqrt{\gamma}/\kappa_0)$ gives a reasonable description of the buckling threshold found in simulations of thermalized shells~\cite{paulose12} with no adjustable parameters. Especially interesting is a result that holds when the pressure difference $p$ between the inside and outside of the shell vanishes, as might be achievable experimentally by creating a hemispherical elastic shell, or a closed shell with regularly spaced large holes. In this case we find that thermal fluctuations must necessarily crush spherical shells larger than a certain temperature-dependent radius given by $R_\textrm{max} = c (\kappa_0/k_B T) \sqrt{\kappa_0/Y_0}$ where the numerical constant $c\approx 160$. Even shells with a small stabilizing outward pressure can be crushed by thermal fluctuations (see Fig.~\ref{fig:buckling_pressure}). We conclude in Sec.~\ref{sec:conclusions} by estimating the importance of thermal fluctuations for a number of thin shells that arise naturally in biology and materials science. For a very thin polycrystalline monolayer shell of a graphene like material (so that it is approximately amorphous), this radius at room temperature is only $160\textrm{nm}$.
 
\section{Elastic energy of deformation}
\label{sec:elastic_energy}
The elastic energy of a deformed thin spherical shell of radius $R_0$ can be estimated with a shallow-shell theory~\cite{sanders63, koiterB}, which considers a small patch of spherical shell that is nearly flat. This may seem a limiting description at first, but as discussed below, the shell response to thermal fluctuations is completely determined by a smaller elastic length scale 
\begin{equation}
\ell_\textrm{el} = \left(\frac{\kappa_0 R_0^2}{Y_0}\right)^{1/4} \sim \sqrt{R_0 h} \ll R_0,
\label{eq:elastic_length}
\end{equation}
where $\kappa_0$ is the microscopic bending rigidity, $Y_0$ is the microscopic Young's modulus and we introduced the effective thickness $h \sim \sqrt{\kappa_0/Y_0}$. For thin shells we require that $h\ll R_0$ or equivalently that the F\"oppl-von Karman number $\gamma=Y_0 R_0^2/\kappa_0 \gg 1$.~\cite{lidmar03}

For a nearly flat patch of spherical shell it is convenient to use the Monge representation near the South Pole to describe the reference undeformed surface
\begin{equation}
{\bf X}_u(x,y) =  x \hat{\bf e}_x +  y \hat{\bf e}_y  + w(x,y) \hat{\bf e}_z,
\end{equation}
where $w(x,y)\approx (x^2+y^2)/(2R_0)$, and then decompose the displacements of a thermally deformed shell configuration ${\bf X}_d(x,y)$ into tangential displacements $u_i (x ,y)$ and radial displacements $f (x ,y)$, such that
\begin{equation}
{\bf X}_d = {\bf X}_u + u_x \hat {\bf t}_x + u_y \hat {\bf t}_y + f \hat {\bf n},
\end{equation}
where $\hat{\bf t}_i = [\hat{\bf e}_i + (\partial_i w) \hat{\bf e}_z]/\sqrt{1 + (\partial_i w)^2}$ is a unit tangent vector, $\hat{\bf n} = [\hat{\bf e}_z - (\partial_i w) \hat{\bf e}_i]/\sqrt{1 + \sum_i (\partial_i w)^2}$ is a unit normal vector that points inward from the South Pole and $i \in \{x,y\}$. Note that positive radial displacements $f(x,y)$ correspond to shrinking of the spherical shell. With this decomposition, the free energy cost of shell deformation can be described as~\cite{koiterB}
\begin{equation}
F = \int \! dxdy \ \left[\frac{\kappa_0}{2} (\nabla^2 f)^2 + \frac{\lambda_0}{2} u_{ii}^2 + \mu_0 u_{ij}^2 - p_0 f \right],
\label{eq:free_energy0}
\end{equation}
where summation over indices $i,j\in\{x,y\}$ is implied. The first term describes the bending energy with a microscopic bending rigidity $\kappa_0$ and next two terms describe the in-plane stretching energy  with two-dimensional Lam\'{e} constants $\lambda_0$ and $\mu_0$; the corresponding Young's modulus is $Y_0=4 \mu_0 (\mu_0+\lambda_0)/(2 \mu_0 + \lambda_0)$. The last term describes the external pressure work, where $p_0$ is a combination of hydrostatic and osmotic contributions. We assume that the interior and exterior of spherical shell is filled with a fluid such as water, which can pass freely through a semipermeable shell membrane on the relevant time scales. Additionally, there may be nonpermeable molecules inside or outside the shell giving rise (within ideal solution theory) to an osmotic pressure contribution $k_B T (c_\textrm{out}-c_\textrm{in})$.~\cite{phillipsB} Here, $c_\textrm{out}$ and $c_\textrm{in}$ are the concentrations of such molecules outside and inside the shell, respectively. Note that for $p_0>0$, introduction of thermal fluctuations into Eq.~(\ref{eq:free_energy0}) requires that we deal with the statistical mechanics of a metastable state -- a macroscopic inversion of the shell (``snap-through'' transition) can lower the free energy,~\cite{landauB} although often with a very large energy barrier.

In the shallow shell approximation the strain tensor is~\cite{koiterB}
\begin{equation}
u_{ij} = \frac{1}{2} (\partial_i u_j + \partial_j u_i) + \frac{1}{2} (\partial_i f)(\partial_j f) - \delta_{ij} \frac{f}{R_0},
\label{eq:strain_tensor}
\end{equation}
where $\delta_{ij}$ is the Kronecker delta. The first term describes the usual linear strains due to tangential displacements. The second describes similar in-plane strains due to displacements in the direction of the surface normals; this nonlinear term makes the analysis of thin plates and shells quite challenging.~\cite{nelsonB} The last term of Eq.~(\ref{eq:strain_tensor}), which linearly couples radial deformations $f(x,y)$ to the sphere curvature $1/R_0$, tells us that spherical shells cannot be bent without stretching, a striking change from flat plates where $R_0 \rightarrow \infty$. The importance of this stretching can be estimated by considering a small radial deformation of amplitude $f_0$ over some characteristic length scale $\ell$, such that the non-linear term $(\partial_i f)(\partial_j f)/2$ in the strain tensor $u_{ij}$ is negligible. The bending energy cost scales as $\sim \kappa_0 f_0^2/\ell^4$, while the stretching energy cost scales as $\sim Y_0 f_0^2/R_0^2$. The bending energy  dominates for deformations on small scales $\ell\ll \ell_\textrm{el}$, while the stretching energy cost dominates for deformations on large scales $\ell \gg \ell_\textrm{el}$, where the transition elastic length scale $\ell_\textrm{el}$ was defined in Eq.~(\ref{eq:elastic_length}).

\section{Thermal fluctuations}
\label{sec:thermal_fluctuations}

The effects of thermal fluctuations are reflected in correlation functions obtained from functional integrals such as~\cite{nelsonB, katsnelsonB, paulose12}
\begin{subequations}
\label{eq:correlations}
\begin{eqnarray}
\left< f_0 \right>  \equiv \left< f({\bf r}_1) \right> &=& \frac{1}{Z} \int \! \mathcal{D}[u_i,f] \ f({\bf r}_1) e^{-F/k_B T}, \\
G_{ff} ({\bf r_2}-{\bf r_1})& \equiv & \left<\delta f({\bf r}_1) \delta f({\bf r}_2) \right>, \nonumber \\
G_{ff} ({\bf r_2}-{\bf r_1})&=& \frac{1}{Z} \int \! \mathcal{D}[u_i,f] \ \delta f({\bf r}_1) \delta f({\bf r}_2) e^{-F/k_B T},\nonumber\\
\\
Z & =& \int \! \mathcal{D}[u_i,f] e^{-F/k_B T},
\end{eqnarray}
\end{subequations}
where $T$ is the ambient temperature, $k_B$ is Boltzmann's constant, ${\bf r}\equiv(x,y)$ and $\delta f({\bf r}) = f({\bf r}) -\left< f_0 \right>$. Here, $f_0$ represents the uniform part of the fluctuating contraction or dilation of the spherical shell. One can define similar correlation functions for tangential displacements $u_i(x,y)$, but they are not the main focus of this study.

Besides separating tangential displacements $u_i({\bf r})$ and radial displacements $f({\bf r})$, it is also useful to further decompose radial displacements as $f({\bf r})=f_0 + \tilde f({\bf r})$, where $f_0$ is the uniform part of the fluctuating radial displacement defined in the above paragraph. The quantity $\tilde f({\bf r})$ is then the deformation with respect to $f_0$, such that $\frac{1}{A}\int\! d^2{\bf r} \, \tilde f=\langle\tilde f\rangle=0$, where $A$ is the area. Finally, it is convenient to integrate out the in-plane phonon degrees of freedom $u_i({\bf r})$ as well as $f_0$ and study the effective free energy for radial displacements. The effective free energy then becomes~\cite{paulose12}
\begin{subequations}
\label{eq:effective_free_energy_both}
\begin{eqnarray}
\label{eq:effective_free_energy_int}
F_\textrm{eff} & = & -k_B T \ln \left(\int \mathcal{D}[u_i, f_0] e^{-F/k_B T} \right),  \\
F_\textrm{eff} & = & \int \! d^2{\bf r} \left(  \frac{1}{2} \left[ \kappa_0 (\nabla^2 \tilde f)^2 - \frac{p_0 R_0}{2} |\nabla \tilde f|^2 + \frac{Y_0 \tilde  f^2}{R_0^2}\right] \right. \nonumber \\
 && \left. + \frac{Y_0}{8} \left[P_{ij}^T (\partial_i \tilde f) (\partial_j \tilde f) \right]^2 - \frac{Y_0 \tilde f}{2 R_0} \left[P_{ij}^T (\partial_i \tilde f) (\partial_j \tilde f) \right] \right),\nonumber \\
\label{eq:effective_free_energy}
\end{eqnarray}
\end{subequations}
where $P_{ij}^T = \delta_{ij}-\partial_i \partial_j/\nabla^2$ is the transverse projection operator. From the effective free energy above, we see that an inward pressure $p_0$ acts like a negative surface tension $\sigma=-p_0 R_0/2$. (A negative outward pressure $p_0<0$ would stabilize the shell, similar to a conventional surface tension.) The two terms that involve both the Young's modulus $Y_0$ and radius $R_0$ are new for spherical shells, and arise from the coupling between radial displacements and in-plane stretching induced by the Gaussian curvature [see Eq.~(\ref{eq:strain_tensor})]. Note that the last term of Eq.~(\ref{eq:effective_free_energy}) breaks the symmetry between inward and outward normal displacements $\tilde f(x,y)$ of the shell.

Functional integrals similar to those in Eqs.~(\ref{eq:correlations}) and Eq.~(\ref{eq:effective_free_energy_int}) determine the average contraction of a spherical shell
\begin{equation}
\left< f_0 \right> = \left< f ({\bf r}_1)\right> = \frac{p_0 R_0^2}{4 (\mu_0 + \lambda_0)} + \frac{R_0}{4} \left<|\nabla \tilde f|^2 \right>,
\label{eq:radius_shrinking}
\end{equation}
where the first term, controlled by the bulk modulus $\mu_0+\lambda_0$, describes the usual mechanical shrinkage due to an inward external pressure $p_0>0$, and the second describes additional contraction due to thermal fluctuations. This additional shrinking arises because nonuniform radial fluctuations $\tilde f({\bf r})$ at fixed radius would increase the integrated area, with a large stretching energy cost. The system prefers to wrinkle and shrink its radius to gain entropy, while keeping the integrated area of the convoluted shell approximately constant.

The effective free energy for radial displacements $\tilde f({\bf r})$ in Eq.~(\ref{eq:effective_free_energy}) suggests that the Fourier transform of the correlation function $G_{ff}({\bf q}) = \int \!(d^2{\bf r}/A) e^{-i {\bf q} \cdot {\bf r}} G_{ff} ({\bf r})$ can be represented as~\cite{paulose12}
\begin{eqnarray}
G_{ff}({\bf q}) &=& \left< |\tilde f({\bf q})|^2 \right>\nonumber \\
G_{ff}({\bf q}) &=& \frac{k_B T}{A \left[\kappa_R(q) q^4 - \frac{1}{2} p_R(q) R_0 q^2 + \frac{Y_R(q)}{R_0^2}\right]},
\label{eq:propagator}
\end{eqnarray}
where $A$ is the area of a patch of spherical shell. The functional form in Eq.~(\ref{eq:propagator}) above is dictated by quadratic terms in Eq.~(\ref{eq:effective_free_energy}); the effect of the anharmonic terms is to replace bare parameters $\kappa_0$, $Y_0$ and $p_0$ with the scale dependent renormalized parameters $\kappa_R(q)$, $Y_R(q)$ and $p_R(q)$ as was shown previously for solid flat membranes in the presence of thermal fluctuations~\cite{nelsonB,katsnelsonB}. Note that the last term in the denominator of Eq.~(\ref{eq:propagator}) suppresses radial fluctuations due to the stretching energy cost and makes them finite even for long wavelength modes (small $q$). Conversely, the amplitude of long wavelength fluctuations diverges more strongly in the limit $R_0\rightarrow \infty$ of large shells.

Before we discuss the renormalizing effect of nonlinearities in Eq.~(\ref{eq:effective_free_energy}), it is useful to note that for large inward external pressure $p_0>0$, the denominator in Eq.~(\ref{eq:propagator}) can become negative for certain wavevectors ${\bf q}$, which indicates that these radial deformation modes $\tilde f ({\bf q})$ become unstable.~\cite{paulose12} If we neglect nonlinear effects, and replace the renormalized couplings $\kappa_R$, $Y_R$ and $p_R$ by their bare values, the minimal value of external pressure $p_c^0$, where these modes first become unstable, is
\begin{equation}
p_c^0 = \frac{4 \sqrt{\kappa_0 Y_0}}{R_0^2},
\label{eq:classical_buckling_pressure}
\end{equation}
which corresponds to the classical buckling pressure for spherical shells~\cite{koiterB}. The magnitude of the wavevectors ${\bf q}_c$ for the unstable modes at the critical external pressure $p_c^0$ is~\cite{hutchinson67}
\begin{equation}
q_c = \left( \frac{Y_0}{\kappa_0 R_0^2}\right)^{1/4} = \ell_\textrm{el}^{-1}.
\end{equation}
When these ideas are extended to finite temperatures, this threshold becomes a limit of metastability, and we expect hysteresis loops as the external pressure is cycled up and down.~\cite{katifori10}

Some insights into the statistical mechanics associated with Eqs.~(\ref{eq:effective_free_energy_int}) and (\ref{eq:effective_free_energy}) follows from calculating the renormalized bending rigidity, Young's modulus and effective pressure at long wavelengths via low temperature perturbation theory in $k_B T/ \kappa_0$. When the external pressure is zero, Paulose \emph{et al.} found that~\cite{paulose12}
\begin{subequations}
\label{eq:perturbation_results}
\begin{eqnarray}
\kappa_R & \approx & \kappa_0 \left[1 + \frac{61}{4096} \frac{k_B T}{\kappa_0} \sqrt{\gamma} \right] , \\
Y_R & \approx & Y_0 \left[1 - \frac{3}{256} \frac{k_B T}{\kappa_0} \sqrt{\gamma} \right] , \\
p_R & \approx & \frac{p_c^0}{24 \pi}  \frac{k_B T}{\kappa_0} \sqrt{\gamma}, 
\label{eq:perturbation_renromalized_pressure}
\end{eqnarray}
\end{subequations}
where $\gamma=Y_0 R_0^2/\kappa_0$ is the F\"{o}ppl-von Karman number and the critical  pressure parameter $p_c^0$ is given by Eq.~(\ref{eq:classical_buckling_pressure}). Perturbation theory reveals that thermal fluctuations enhance the bending rigidity and soften the Young's modulus. However, the corrections to $\kappa_R$ and $Y_R$ are multiplied by $\sqrt{\gamma}$, which \emph{diverges} as the radius $R_0$ of the thermalized sphere tends to infinity. Especially striking is a similar divergence in the effective pressure $p_R$, see Eq.~(\ref{eq:perturbation_renromalized_pressure}). Evidently, even if the microscopic pressure difference $p_0$ between the inside and outside of sphere is zero, thermal fluctuations will nevertheless generate an effective pressure that eventually exceeds the buckling instability of the sphere for sufficiently large $R_0$. A naive estimate for the critical radius $R_\textrm{max}$ can be obtained by requiring that the renormalized pressure $p_R$ becomes equal to the buckling pressure $p_c^0$ in Eq.~(\ref{eq:perturbation_renromalized_pressure}), which leads to $R_\textrm{max} \approx c [\kappa_0/k_B T] \sqrt{\kappa_0/Y_0}$ with $c=24\pi \approx 75$. Some evidence in this direction already appears in the computer simulations of Ref.~\cite{paulose12}, where amorphous thermalized spheres already begin to collapse at less than half the classical buckling pressure (see also Fig.~\ref{fig:simulations}, where the pressure is $36\%$ of $p_c^0$). Similar perturbative divergences in the bending rigidity and Young's modulus of flat membranes of size $R_0$ (here the corrections diverge with $\gamma$ rather than $\sqrt{\gamma}$~\cite{nelsonB}) can be handled with integral equation methods,~\cite{nelson87,ledoussal92} which sum contributions to all orders in perturbation theory, or alternatively, with the renormalization group.~\cite{aronovitz88} It is this latter approach we take in the next Section.

\section{Perturbative renormalization group}
\label{sec:renormalization_group}
The effect of the anharmonic terms in Eq.~(\ref{eq:effective_free_energy}) at a given scale $\ell^*\equiv \pi/q^*$ can be obtained by systematically integrating out all degrees of freedom on smaller scales (i.e., larger wavevectors). Formally this renormalization group transformation proceeds by splitting radial displacements $\tilde f({\bf r})$ into slow modes $\tilde f_<({\bf r}) = \sum_{|\bf {q}|<q^*} e^{i {\bf q} \cdot {\bf r}} \tilde f({\bf q})$ and fast modes $\tilde f_>({\bf r}) = \sum_{|\bf {q}|>q^*} e^{i {\bf q} \cdot {\bf r}} \tilde f({\bf q})$, which are then integrated out as
\begin{equation}
F_\textrm{eff} (\ell^*) =  -k_B T \ln \left(\int \!\mathcal{D}[\tilde f_>] \ e^{-F_\textrm{eff}/k_B T} \right).
\end{equation}
These functional integrals can be approximately evaluated with standard perturbative renormalization group calculations~\cite{amitB} and lead to an effective free energy with the same form as in Eq.~(\ref{eq:effective_free_energy}), except that renormalized parameters become scale dependent, i.e. they are replaced by $\kappa_R(\ell^*)$, $Y_R(\ell^*)$ and $p_R(\ell^*)$.

\begin{figure}[t]
\includegraphics[scale=.5]{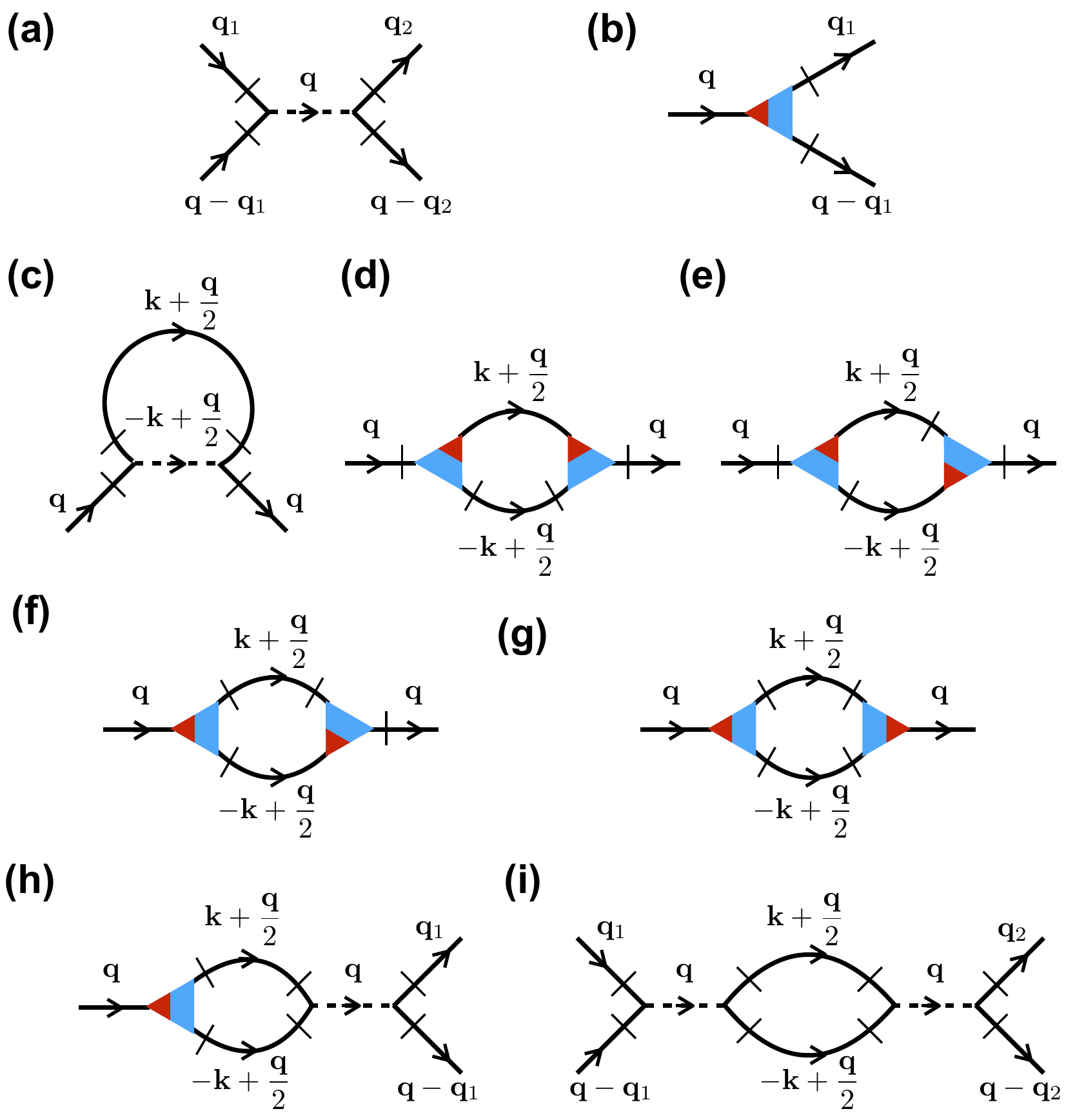}
\caption{(Color online) Feynman diagrams contributing to the renormalization flows of Eq.~(\ref{eq:flow}). (a) Four-point and (b) three-point vertices describe the quartic and cubic terms in the free energy Eq.~(\ref{eq:effective_free_energy}). Legs represent radial displacement fields $\tilde f({\bf q})$ and slashes on legs correspond to spatial derivatives, which lead to additional factors of wave vectors in the Fourier space. The red part of the three-point vertex in (b) connects to a field without a slash, while the blue parts connect to derivative terms. The four-point vertex carries a factor $Y$, while the three-point vertex carries a factor $Y/R$ (c-i) One-loop diagrams that contribute to the renormalization flows of (c-g) the bending rigidity $\kappa_R$, (f-g) the external pressure $p_R$, and (g) the Young's modulus $Y_R$ in the propagator $G_{ f f}({\bf q})$ in Eq.~(\ref{eq:propagator}). Diagrams (h) and (i) describe one-loop contributions to the renormalization flows of the Young's modulus $Y_R$ associated with three-point and four-point vertices, respectively. The connected legs in these diagrams represent the propagators $G_{ f f}({\bf q})$, with wave vectors ${\bf k}$ restricted to the momentum shell $\Lambda/b < k < \Lambda$.
}
\label{fig:diagrams}
\end{figure}

To implement this momentum shell renormalization group, we first integrate out all Fourier modes in a thin momentum shell $\Lambda/b < q < \Lambda$, where $a=\pi/\Lambda$ is a microscopic cutoff (e.g. the shell thickness) and $b\equiv \ell/a=e^{s}$ with $s \ll 1$. Next we rescale lengths and fields~\cite{aronovitz88, radzihovsky91}
\begin{subequations}
\label{eq:rescaling}
\begin{eqnarray}
{\bf r} &=& b {\bf r}', \\
\tilde f({\bf r}) & = & b^{\zeta_f} \tilde f'({\bf r}'),
\end{eqnarray}
\end{subequations}
where the field rescaling exponent $\zeta_f$ will be chosen to simplify the resulting renormalization group equations. We find it convenient to work directly with a $D=2$ dimensional spherical shell embedded in $d=3$ space, rather than introducing an expansion in $\epsilon = 4 -D$.~\cite{aronovitz88} 
Finally, we define new elastic constants $\kappa'$, $Y'$, and a new external pressure $p'$, such that the free energy functional in Eq.~(\ref{eq:effective_free_energy}) retains the same form after the first two renormalization group steps. It is common to introduce $\beta$ functions~\cite{amitB}, which define the renormalization flow of elastic constants. It is not possible to calculate these $\beta$ functions exactly, but one can use diagrammatic techniques~\cite{amitB} to obtain systematic approximations in the limit $s \ll 1$.
To one loop order (see Fig.~\ref{fig:diagrams}) the renormalization group flows are given by
\begin{subequations}
 \label{eq:flow}
 \begin{eqnarray}
 \beta_\kappa  &=&  \frac{d \kappa'}{d s} = 2 (\zeta_f - 1) \kappa' + \frac{3 k_B T Y' \Lambda^2}{16 \pi \mathcal{D}} \nonumber \\ 
 && \quad \quad \quad  - \frac{3 k_B T Y'^2 \Lambda^2}{8 \pi R'^2 \mathcal{D}^2} \bigg[1 +\frac{I_{\kappa1}}{\mathcal{D}^2} + \frac{I_{\kappa2}}{\mathcal{D}^4} \bigg],\\
 \beta_Y &=&  \frac{d Y'}{d s} = 2 \zeta_f Y' - \frac{3 k_B T Y'^2 \Lambda^6}{32 \pi  \mathcal{D}^2},  \label{eq:renormalization_flow_Y} \\
 \beta_{p}  &=&  \frac{d p'}{d s}= (2 \zeta_f +1) p' + \frac{3 k_B T Y'^2 \Lambda^4}{4 \pi R'^3 \mathcal{D}^2} \left[1 + \frac{I_p}{\mathcal{D}^2} \right], \quad \  \label{eq:flow_p} \\
 \beta_{R}  &=&  \frac{d R'}{d s} = - R',
 \end{eqnarray} 
 \end{subequations}
 where we introduced the denominator term
  \begin{eqnarray}
  \mathcal{D}&=& \kappa' \Lambda^4 - \frac{p' R' \Lambda^2}{2} + \frac{Y'}{R'^2}.
  \end{eqnarray} 
The derivation of recursion relations in Eq.~(\ref{eq:flow}) is given in the Appendix~\ref{sec:app_rg_flow}, where we also provide detailed expressions for $I_{\kappa1}$, $I_{\kappa2}$ and $I_p$ in Eq.~(\ref{eq:app:rg_flow_detailed}).
The $\beta_Y$ recursion relation in Eq.~(\ref{eq:renormalization_flow_Y}) describes changes in the quadratic ``mass'' proportional to $Y$ in Eq.~(\ref{eq:effective_free_energy}). Similarly, we can calculate the recursion relations for the cubic and quartic terms in Eq.~(\ref{eq:effective_free_energy}) and we find that the only significant change is that the $2 \zeta_f Y$ term now becomes $(3 \zeta_f - 1) Y$ and $(4 \zeta_f -2) Y$, respectively. To ensure that the free energy retains the same form after the first two steps in the renormalization procedure, we choose $\zeta_f=1$, so that these three terms renormalize in tandem. The final results are independent of the precise choice of $\zeta_f$, as illustrated in Appendix~\ref{sec:app:rg_flow_sheets} for thermalized flat sheets.

The scale-dependent parameters $\kappa'(s)$, $Y'(s)$, and $p'(s)$, obtained by integrating the differential equations in Eqs.~(\ref{eq:flow}) up to a scale $s=\ln(\ell/a)$ with initial conditions $\kappa'(0)=\kappa_0$, $Y'(0)=Y_0$ and $p'(0)=p_0$, are related to the scaling of propagator $G_{ff}(q)$ as~\cite{amitB}
\begin{eqnarray}
\!\!&&\!\!G_{ff}(q|\kappa_0, p_0, Y_0, R_0, A) = \left< |f({\bf q})|^2\right> = e^{2 \zeta_f s} \left< |f'({\bf q}')|^2\right>\nonumber \\
\!\!&&\!\!= e^{2 \zeta_f s} G_{ff}(q e^s | \kappa'(s), p'(s), Y'(s), R_0 e^{-s}, A e^{-2 s}),
\label{eq:scaling_relation}
\end{eqnarray}
where we explicitly insert the rescaled momenta $q'=q e^s$, the rescaled radius $R'=R_0 e^{-s}$ and the rescaled patch area $A'=Ae^{-2 s}$. By replacing the left hand side in the Eq.~(\ref{eq:scaling_relation})  above with the renormalized propagator $G_{ff}(q)$ in Eq.(\ref{eq:propagator}), we find the scale-dependent renormalized parameters
\begin{subequations}
\begin{eqnarray}
\kappa_R (s) & = & \kappa' (s) e^{(2-2 \zeta_f) s} = \kappa' (s),  \\
Y_R (s) & = & Y' (s) e^{(-2 \zeta_f) s} = Y' (s) e^{-2 s},  \\
p_R (s) & = & p' (s) e^{(-1-2 \zeta_f) s}= p' (s) e^{-3 s}, 
\end{eqnarray}
\end{subequations}
where we used $\zeta_f=1$ and parameter $s$ is related to the length scale $\ell = a e^s$ or equivalently to the magnitude of wavevector $q\equiv \pi/\ell$.

Note that by sending the shell radius to infinity ($R_0\rightarrow \infty$) and the pressure $p_0\rightarrow 0$, such that the product $\sigma=-p_0 R_0/2$ remains fixed in Eq.~(\ref{eq:effective_free_energy}), we recover the renormalization flows for solid flat membranes with the addition of a tension $\sigma$.~\cite{radzihovsky91,kosmrlj16} However, for spherical shells with finite $R_0$ thermal fluctuations renormalize and effectively increase the external pressure [see Eq.~(\ref{eq:flow_p})], in striking contrast to the behavior of flat membranes. Note, in particular,  that an effective pressure is generated by Eq.~(\ref{eq:flow_p}), even if the microscopic pressure $p_0$ vanishes!

Before discussing the detailed renormalization group predictions for spherical shells, it is useful to recall that for flat membranes with no tension, thermal fluctuations become important on scales larger than thermal length~\cite{nelson87,aronovitz88,radzihovsky91,kosmrlj16,guitter89}
\begin{equation}
\ell_\textrm{th} = \sqrt{\frac{16 \pi^3 \kappa_0^2}{3 k_B T Y_0}},
\label{eq:thermal_length}
\end{equation}
and the renormalized elastic constants become strongly scale-dependent,
\begin{eqnarray}
\kappa_R(\ell)&\sim& \left\{ 
\begin{array}{c l}
\kappa_0,  & \ell \ll \ell_\textrm{th} \\
\kappa_0 (\ell / \ell_\textrm{th})^{\eta}, \quad & \ell_\textrm{th} \ll \ell 
\end{array}
\right. , \nonumber \\
Y_R(\ell)&\sim& \left\{ 
\begin{array}{c l}
Y_0 , & \ell \ll \ell_\textrm{th} \\
Y_0 (\ell/\ell_\textrm{th})^{-\eta_u}, & \ell_\textrm{th} \ll \ell\\
\end{array}
\right. ,
\label{eq:renormalized_elastic_constants}
\end{eqnarray}
where $\eta\approx 0.80$-$0.85$~\cite{nelson87, aronovitz88, radzihovsky91, guitter89, ledoussal92,kosmrlj16} and the exponents $\eta$ and $\eta_u$ are connected via a Ward identity $\eta_u + 2 \eta=2$ associated with rotational invariance.~\cite{aronovitz88,guitter89} In the one-loop approximation used here for 2d membranes embedded in three dimensions we obtain~\cite{kosmrlj16} $\eta=0.80$, which is adequate for our purposes. In the absence of an external tension, the renormalized bending rigidity $\kappa_R$ can become very large and the renormalized Young's modulus $Y_R$ can become very small for large solid membranes in the flat phase, as seems to be the case for graphene.~\cite{blees15,nicholl15} However, positive external tension acts as an infrared cutoff and the renormalized constants remain finite beyond a tension-induced length scale.~\cite{kosmrlj16,roldan11}

Although the scaling relation $\eta_u + 2 \eta = 2$ originally arose from a Ward identity, \cite{aronovitz88,guitter89} an alternative derivation provides additional physical insight: Suppose we are given a two-dimensional material (graphene, MoS$_2$, the spectrin skeleton of red blood cells, etc.) with a 2d Young's modulus $Y_0$ and a 2d bending rigidity $\kappa_0$. With these material parameters we associate the elastic constants of an equivalent isotropic bulk material with 3d Young's modulus $E_0$, 3d Poisson's ratio $\nu_0$ and thickness $h$ by~\cite{landauB}
\begin{equation}
\kappa_0 = \frac{E_0 h^3}{12 (1-\nu_0^2)}, \quad Y_0= E_0 h.
\label{eq:elastic_constants_2d_3d}
\end{equation}
When thermal fluctuations are considered, we obtain the scale-dependent, 2d elastic parameters displayed in Eq.~(\ref{eq:renormalized_elastic_constants}), $\kappa_R(\ell)\approx \kappa_0 (\ell/\ell_\textrm{th})^\eta$ and $Y_R(\ell) \approx Y_0 (\ell/\ell_\textrm{th})^{-\eta_u}$, where $\ell_\textrm{th} \ll \ell \ll L$, $L$ is the system size and the corresponding scale-dependent 2d Poisson's ratio $\nu(\ell)$ remains of order unity.~\cite{aronovitz88} From these results and equation (\ref{eq:elastic_constants_2d_3d}) we can define a scale-dependent effective thickness $h^2_\textrm{eff}(\ell) \sim \kappa_R (\ell)/Y_R(\ell)$, so  that
\begin{equation}
h^2_\textrm{eff} (\ell) \sim h^2 \left(\ell/\ell_\textrm{th}\right)^{\eta+\eta_u}.
\label{eq:height_fluctuations_1}
\end{equation}
For a $10\mu\textrm{m} \times 10\mu\textrm{m}$ square graphene, where $\ell_\textrm{th}\approx 1 nm$ at room temperature, this thermal amplification (assuming $\eta + \eta_u \approx 0.8 + 0.4=1.2$) converts an atomic thickness to an effective thickness, whose ratio to the size of graphene sheet matches that of the ordinary writing paper, suggesting that room temperature graphene ribbons and springs can be studied with simple paper models.~\cite{blees15} To determine a scaling relation between $\eta$ and $\eta_u$, we note that an alternative definition of the effective thickness follows from~\cite{nelsonB}
\begin{eqnarray}
h^2_\textrm{eff} (\ell) & = & \left< f({\bf r})^2)\right>_\ell \nonumber \\
h^2_\textrm{eff} (\ell) & = &  \int_{|\bf q| \geq \pi/\ell} \frac{d^2 {\bf q}}{(2 \pi)^2} \frac{k_B T}{\kappa_R(q) q^4} \sim \ell^{2-\eta},
\label{eq:height_fluctuations_2}
\end{eqnarray}
where the average is evaluated over a $\ell\times\ell$ patch of the membrane, so that $q\geq \pi/\ell$ in the integration. Requiring similar scaling of Eqs.~(\ref{eq:height_fluctuations_1}) and (\ref{eq:height_fluctuations_2}) with $\ell$ leads to $\eta_u + 2 \eta = 2$.

By rewriting the renormalization group flows in Eq.~(\ref{eq:flow}) in dimensionless form it is easy to see that the renormalized parameters can be expressed in terms of the following scaling functions of dimensionless important length scales and of $p_0/p_c^0$, where $p_c^0$ is the classical buckling pressure in Eq.~(\ref{eq:classical_buckling_pressure}).

\begin{figure*}[t]
\includegraphics[scale=.5]{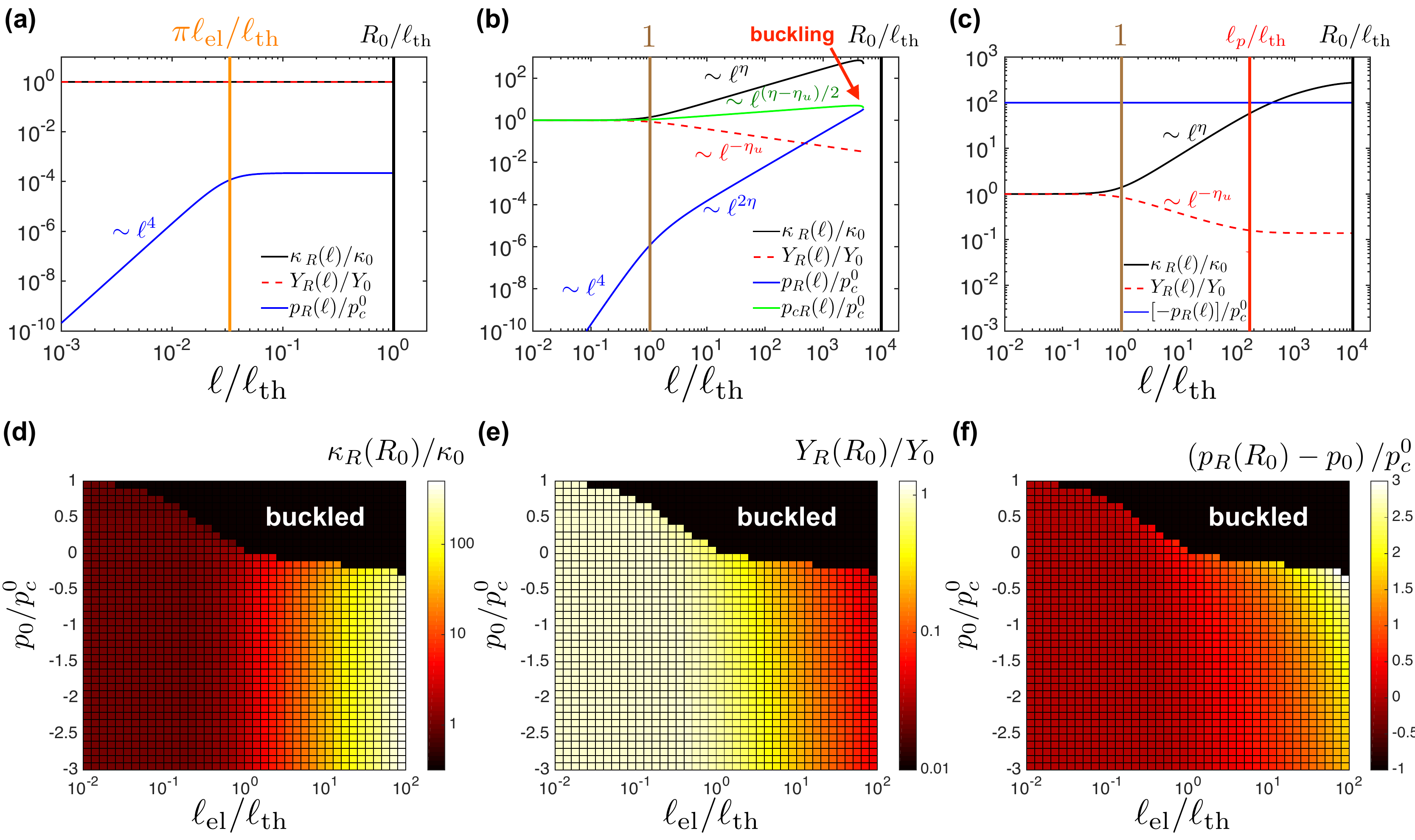}
\caption{(Color online) Typical renormalization group flows at various microscopic pressures $p_0$ and ratios of $\ell_\textrm{el}=(\kappa_0 R_0^2/Y_0)^{1/4}$ to $\ell_\textrm{th}=\sqrt{16 \pi^3 \kappa_0/(3 k_B T Y_0)}$. (a) Spherical shells at low temperature with $\ell_\textrm{el}/\ell_\textrm{th}=10^{-2}$, $R_0/\ell_\textrm{th}=1$, $a/R_0=10^{-6}$ and zero external pressure ($p_0=0$). In this case, there is practically no renormalization of the elastic constants $\kappa_R$ and $Y_R$, while the renormalization of the thermally generated external pressure $p_R$ is cut off at $\ell \approx \pi \ell_\textrm{el}$. (b-c) Spherical shells at high temperature with $\ell_\textrm{el}/\ell_\textrm{th}=10^{2}$, $R_0/\ell_\textrm{th}=10^4$, $a/R_0=10^{-6}$, and (b) zero external pressure ($p_0=0$) or (c) large stabilizing outward pressure ($p_0/p_c = -100$). In both these cases the elastic constants $\kappa_R$ and $Y_R$ initially renormalize in the same way as flat membranes [see Eq.~(\ref{eq:renormalized_elastic_constants})]. In case (b), even in the absence of external pressure $p_0=0$, this large shell buckles, because the thermally generated inward pressure $p_R(\ell)$ eventually reaches the renormalized critical buckling pressure $p_{cR}(\ell)$. In case (c) with a large outward pressure ($p_0<0$), spherical shells remain stable and the renormalization of elastic constants is cut off at the scale $\ell_p/\ell_\textrm{th} \sim(p_c^0/|p_0|)^{1/(2-\eta)} (\ell_\textrm{el}/\ell_\textrm{th})^{2/(2-\eta)} \sim \left(k_B T Y_0/|p_0| R_0 \kappa_0\right)^{1/(2-\eta)}$, which is analogous to the cut off prowided by an outward in-plane tension in flat solid membranes~\cite{kosmrlj16}. For sufficiently large internal pressure $p_0 \lesssim  - k_B T Y_0/ R_0 \kappa_0$ (not shown) the renormalization of $\kappa$ and $Y$ is completely suppressed. 
(d-f) Heat maps of (d) the renormalized bending rigidity $\kappa_R(R_0)$, (e) the renormalized Young's modulus $Y_R(R_0)$, and (f) the thermally induced part of renormalized external pressure $p_R(R_0)-p_0$ evaluated at the scale of the shell radius $R_0$. In (d-f) we used $R_0/\ell_\textrm{el}=10^2$ and $a/R_0=10^{-6}$. The large black buckled region is a direct consequence of thermal fluctuations. Note that both positive (inward) and negative (outward) pressures appear along the y-axis.
}
\label{fig:RGflows}
\end{figure*}

\begin{subequations}
\label{eq:scaling_functions}
\begin{eqnarray}
\kappa_R (\ell) & = & \kappa_0 \,\Phi_\kappa \left(\frac{\ell}{\ell_\textrm{th}}, \frac{\ell_\textrm{el}}{\ell_\textrm{th}},\frac{p_0}{p_c^0},\frac{a}{\ell_\textrm{th}} \right),  \\
Y_R (\ell) & = & Y_0 \,\Phi_Y \left(\frac{\ell}{\ell_\textrm{th}}, \frac{\ell_\textrm{el}}{\ell_\textrm{th}},\frac{p_0}{p_c^0},\frac{a}{\ell_\textrm{th}} \right),  \\
p_R (\ell) & = & p_c^0 \,\Phi_p \left(\frac{\ell}{\ell_\textrm{th}}, \frac{\ell_\textrm{el}}{\ell_\textrm{th}},\frac{p_0}{p_c^0},\frac{a}{\ell_\textrm{th}} \right).
\end{eqnarray}
\end{subequations}
We expect that the scaling functions above are insensitive to the choice of  microscopic cutoff $a$ (e.g. shell thickness or a carbon-carbon spacing in a large spherical buckyball), provided this cutoff is much smaller than other relevant lengths ($a \ll \ell_\textrm{th}, \ell_\textrm{el}$). In principle, we could evaluate renormalized parameters on the whole interval $\ell \in [a,R]$, but for some values of bare parameters $\kappa_0$, $Y_0$, $p_0$ the renormalization flows in Eq.~(\ref{eq:flow}) diverge, when denominators become zero. This singularity indicates the buckling of thermalized spherical shells, which occurs when the renormalized external pressure $p_R(\ell^*)$ reaches the renormalized critical buckling pressure
\begin{equation}
p_{cR}(\ell^*) \equiv \frac{4\sqrt{\kappa_R(\ell^*) Y_R(\ell^*)}}{R_0^2},
\end{equation}
where $\ell^*$ corresponds to the length scale of the unstable mode. In order for the shell to remain stable in the presence of thermal fluctuations, the renormalized pressure $p_R(\ell)$ has to remain below the renormalized critical buckling pressure $p_{cR}(\ell)$ for every $\ell \in [a,R_0]$.

Fig.~\ref{fig:RGflows} displays some typical flows of renormalized parameters. We find that for spherical shells the renormalized elastic constants, initially renormalize in the same way as for  flat membranes [see Eq.~(\ref{eq:renormalized_elastic_constants})], but these singularities are eventually cut off by the Gaussian curvature. At low temperatures ($\ell_\textrm{el} / \ell_\textrm{th} \propto \sqrt{k_B T/\kappa_0} \gamma^{1/4} \ll 1$) and small inward pressures $p_0$, the corrections to renormalized bending rigidity $\kappa_R(\ell)/\kappa_0$ and renormalized Young's modulus $Y_R(\ell)/Y_0$  grow as $(k_B T/\kappa_0) Y_0 \ell^2/\kappa_0$, while the renormalized pressure $p_R(\ell)-p_0$ grows as $k_B T Y_0^2 \ell^4/(\kappa_0^2 R_0^3)$. The renormalization is cut off at the elastic length scale $\ell_\textrm{el}$ (see Fig.~\ref{fig:RGflows}a), where the $Y'/R'^2$ term starts dominating over the $\kappa' \Lambda^4$ term in denominators $\mathcal{D}$ of the recursion relations in Eqs.~(\ref{eq:flow}). This cutoff gives rise to corrections of size $(k_B T/\kappa_0) \sqrt{Y_0 R_0^2/\kappa_0}$ [see Eq.~(\ref{eq:perturbation_results})] for spherical shells, in contrast to the corrections of size $(k_B T/\kappa_0) Y_0 L_0^2/\kappa_0$ for flat sheets of size $L_0$.

At high temperatures ($\ell_\textrm{el} / \ell_\textrm{th} \propto \sqrt{k_B T/\kappa_0} \gamma^{1/4} \gg 1$) and small external pressures $p_0$, the corrections to the renormalized parameters $\kappa_R(\ell)$, $Y_R(\ell)$, $p_R(\ell)$ initially still grow in the same way as described above for low temperatures. However, a transition to the new regime happens at the thermal length scale $\ell_\textrm{th} \sim \kappa_0/\sqrt{k_B T Y_0}$, where corrections to the renormalized bending rigidity $\kappa_R(\ell_\textrm{th})/\kappa_0$ and the renormalized Young's modulus $Y_R(\ell_\textrm{th})/Y_0$ become of order unity and the renormalized pressure is $p_R(\ell_\textrm{th})-p_0 \sim p_c^0 (\ell_\textrm{th}/\ell_\textrm{el})^2 \ll p_c^0$. On scales larger than the thermal length scale the renormalized parameters scale according to
\begin{subequations}
\begin{eqnarray}
\kappa_R(\ell) &\sim& \kappa_0 (\ell/\ell_\textrm{th})^\eta,\\
Y_R(\ell) &\sim& Y_0 (\ell/\ell_\textrm{th})^{-\eta_u},\\
p_R(\ell)-p_0 &\sim& p_c^0 (\ell_\textrm{th}/\ell_\textrm{el})^2 (\ell/\ell_\textrm{th})^{2 \eta},
\end{eqnarray}
\end{subequations}
where $\eta=0.8$ and $\eta_u=0.4$ are the same exponents as for flat sheets. If the external pressure $p_0$ is properly tuned, such that the renormalized pressure $p_R(\ell)$ remains small, then the renormalization gets cut off at the length scale $\ell^*$, where the $Y'/R'^2$ term starts dominating over the $\kappa' \Lambda^4$ term in denominators of recursion relations in Eqs.~(\ref{eq:flow}). This scale is given by
\begin{equation}
\ell^* \sim \ell_\textrm{th} \left(\frac{\ell_\textrm{el}}{\ell_\textrm{th}} \right)^{4/(4-\eta-\eta_u)} \!\sim  \ell_\textrm{th} \left(\frac{\ell_\textrm{el}}{\ell_\textrm{th}} \right)^{4/(2+\eta)} \!\propto R_0^{2/(2+\eta)},
\label{eq:lstar}
\end{equation}
where we used the exponent relation $\eta_u+2 \eta=2$. Due to this cutoff we now find renormalized bending rigidity $\kappa_R(R_0) \propto R_0^{2 \eta/(2+\eta)}$ and the renormalized Young's modulus $Y_R(R_0) \propto R_0^{-2 \eta_u/(2+\eta)}$, which is again different from flat sheets of size $L$ ($\kappa_R(L) \propto L^\eta$, $Y_R(L)\propto L^{-\eta_u}$). Note that in the absence of a microscopic pressure ($p_0\equiv 0$) thermal fluctuations generate a renormalized pressure $p_R(\ell^*) \sim p_c^0 \left[\ell_\textrm{el}/\ell_\textrm{th}\right]^{(6 \eta-4)/(2+\eta)}$, which is of the same order as the renormalized buckling pressure $p_{cR}(\ell^*)=4 \sqrt{\kappa_R(\ell^*) Y_R(\ell^*)}/R_0^2 \sim p_c^0 \left[\ell_\textrm{el}/\ell_\textrm{th}\right]^{(6 \eta-4)/(2+\eta)}$. Numerically we find that at zero external pressure the renormalized pressure $p_R(\ell^*)$ is actually large enough to crush the shell (see Fig.~\ref{fig:RGflows}b). In fact, spherical shells can only be stable if the outward pressure is larger than
\begin{eqnarray}
p_{0,\textrm{min}} &=& -\mathcal{C}_1 p_c^0 \left(\frac{\ell_\textrm{el}}{\ell_\textrm{th}}\right)^{(6 \eta-4)/(2+\eta)}, \nonumber \\
 &=& -\mathcal{C}_2  p_c^0 \left(\frac{k_B T}{\kappa_0} \sqrt{\frac{Y_0 R_0^2}{\kappa_0}}\right)^{(3 \eta-2)/(2+\eta)},
\label{eq:min_p}
\end{eqnarray}
where we find $C_1 \approx 0.10$, $C_2 \approx 0.047$ and $(3\eta-2)/(2+\eta)\approx 0.14$. For large outward pressures ($p_0 \ll p_{0,\textrm{min}}<0$) the renormalization gets cut off at a pressure length scale $\ell_p$ given by
\begin{equation}
\frac{\ell_p}{\ell_\textrm{th}} \sim\left(\frac{p_c^0}{|p_0|}\right)^{1/(2-\eta)} \left(\frac{\ell_\textrm{el}}{\ell_\textrm{th}}\right)^{2/(2-\eta)} \sim \left(\frac{k_B T Y_0}{|p_0| R_0 \kappa_0}\right)^{1/(2-\eta)},
\end{equation}
when the $p' R' \Lambda^2$ term starts dominating over the $\kappa' \Lambda^4$ and $Y'/R'^2$ terms in denominators of recursion relations in Eq.~(\ref{eq:flow}). As can be seen from Fig.~\ref{fig:RGflows}c, the Young's modulus $Y_R(\ell)$  stops renormalizing at the length scale $\ell_p$, while the renormalization of bending rigidity still continues until the $Y'/R'^2$ term in denominators of recursion relations in Eq.~(\ref{eq:flow}) starts to dominate. Note that for sufficiently large internal pressure $p_0 \ll - k_B T Y_0/(R_0 \kappa_0)$, the cut off length scale $l_p$ becomes smaller than the thermal length scale $\ell_\textrm{th}$ and the effects of thermal fluctuations are completely suppressed.

In Fig.~\ref{fig:RGflows} we also present heat maps of (d) the renormalized bending rigidity $\kappa_R(R_0)$, (e) the renormalized Young's modulus $Y_R(R_0)$, and (f) the thermally induced part of renormalized external pressure $p_R(R_0)-p_0$ evaluated at the scale of shell radius $R_0$, as a function of $p_0/p_c^0$ and $\ell_\textrm{el}/\ell_\textrm{th} \propto \sqrt{k_B T /\kappa_0}\, \gamma^{1/4}$. These are the renormalized parameters that one could measure in experiments by analyzing the long wavelength radial fluctuations described by Eq.~(\ref{eq:propagator}), once the thermal fluctuations are cut off by either the elastic length ($\ell_\textrm{el}$) or a sufficiently large outward pressure ($p_0<0$), which stabilizes the shells. Although the scaling functions in Eq.~(\ref{eq:scaling_functions}) could in principle depend directly on the shell size $R_0$, this is not the case, because the renormalization group cutoffs at $\ell_p$ or $\ell_\textrm{el}$ intervene before $\ell=R_0$.

\begin{figure}[t]
\includegraphics[scale=.5]{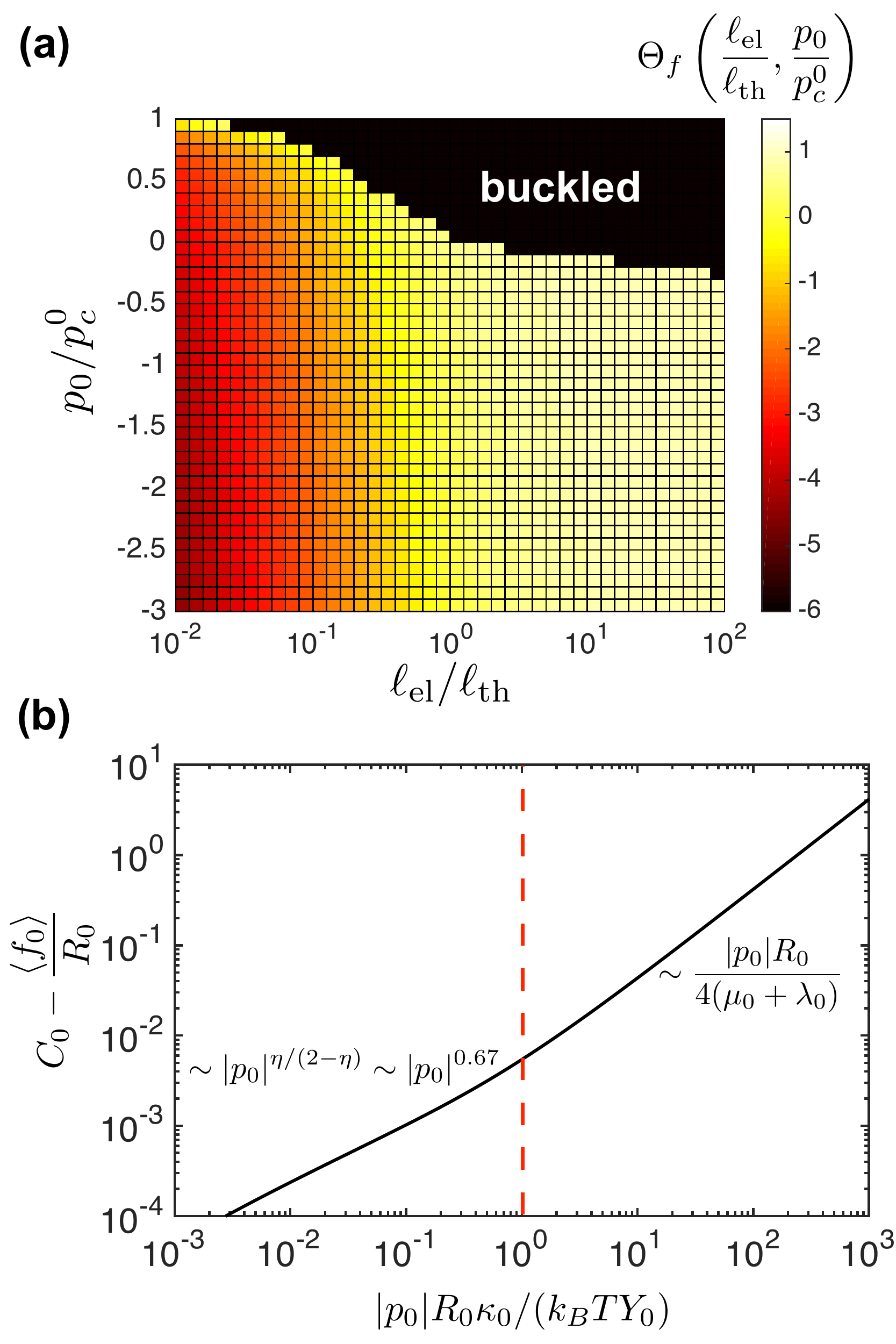}
\caption{(Color online) Heat map depicting the average thermal shrinking of the shell radius $\left<f_0 \right>$, as described by the scaling function $\Theta_f \left(\ell_\textrm{el}/\ell_\textrm{th}, p_0/p_c^0 \right)$ [see Eq.~(\ref{eq:scaling_theta})].  (a) Contours of  the scaling function $\Theta_f$  are shown with $R_0/\ell_\textrm{el}=10^2$, $a/R_0=10^{-6}$. (b) Non-linear response for large membranes ($\ell_\textrm{th}\ll \ell_\textrm{el}$) under large outward pressure $p_0<0$ [see  Eq.~(\ref{eq:shrinking_nonlinear})]. Here, the parameter on the $y$-axis $C_0 \approx  k_B T/(8 \pi \kappa_0)\left[ \ln\left(\ell_\textrm{th}/a\right) + 1/\eta\right]$, whereas $\ell_\textrm{el}/\ell_\textrm{th}=10^1$, $R_0/\ell_\textrm{el}=10^3$, $a/R_0=10^{-6}$.
}
\label{fig:radius_shrinking}
\end{figure}

In experiments one could also measure the average thermal shrinking of the shell radius $\left<f_0\right>$ [see Eq.~(\ref{eq:radius_shrinking})], relative to its $T=0$ value, which is related to the integral of the correlation functions in Eq.~(\ref{eq:propagator}),
\begin{eqnarray}
\left<f_0 \right> &\approx& \frac{p_0 R_0^2}{4 (\mu_0 + \lambda_0)} + \frac{R_0}{8 \pi} \int_{\pi/R}^{\pi/a} \! dq \ q^3 G_{ f  f}(q) A, \nonumber \\
\left<f_0 \right> &\equiv& \frac{p_0 R^2}{4 (\mu_0 + \lambda_0)} + \frac{k_B T R_0}{8 \pi \kappa_0} \Phi_f \left(\frac{R_0}{\ell_\textrm{th}}, \frac{\ell_\textrm{el}}{\ell_\textrm{th}},\frac{p_0}{p_c^0},\frac{a}{\ell_\textrm{th}} \right).\nonumber \\
\label{eq:radius_shrinking_scaling}
\end{eqnarray}
Here, $A$ is the area of the patch that defines shallow shell theory; it drops out of the scaling function defined by the second line -- see Eq.~(\ref{eq:propagator}).
Note that the integral above diverges logarithmically for $q \lesssim \pi/a$, i.e. at distances close to the microscopic cutoff $a$, where $G_{ff}(q) \approx k_B T/(A\kappa_0 q^4)$. This divergent part can be subtracted from the scaling function $\Phi_f$ defined in the second part of Eq.~(\ref{eq:radius_shrinking_scaling});  the remaining piece, which we call $\Theta_f$, is approximately independent of the microscopic cutoff $a$ and the shell size $R_0$. Fig.~\ref{fig:radius_shrinking}a shows via a heat map how the scaling function $\Theta_f$ depends on the other important parameters, $\ell_\textrm{el}/\ell_\textrm{th}\propto \sqrt{k_B T/\kappa_0} \, \gamma^{1/4}$ and $p_0/p_c^0$. The average shrinking of the shell radius can then be expressed as
\begin{equation}
\left<f_0 \right> = \frac{p_0 R_0^2}{4 (\mu_0 + \lambda_0)} + \frac{k_B T R_0}{8 \pi \kappa_0}\left[ \ln\left(\frac{\ell_\textrm{th}}{a}\right) + \Theta_f \left( \frac{\ell_\textrm{el}}{\ell_\textrm{th}},\frac{p_0}{p_c^0} \right)\right].
\label{eq:scaling_theta}
\end{equation}

Finally, we find that for large shells with $\ell_\textrm{th} \ll \ell_\textrm{el}$ that are under a stabilizing outward pressure ($p_0<0$), the renormalization procedure leads to a nonlinear dependence of the average shell radius shrinkage $\left<f_0\right>$ with internal pressure $|p_0|$ as (see Fig.~\ref{fig:radius_shrinking}b) 
\begin{eqnarray}
\left<f_0 \right> &\approx& -\frac{|p_0| R_0^2}{4 (\mu_0 + \lambda_0)} + \frac{k_B T R_0}{8 \pi \kappa_0}\left[ \ln\left(\frac{\ell_\textrm{th}}{a}\right) + \frac{1}{\eta}\right]\nonumber \\
&& -\mathcal{C} \frac{k_B T R_0}{\kappa_0} \left(\frac{|p_0| R_0 \kappa_0}{k_B T Y_0} \right)^{\eta/(2-\eta)},
\label{eq:shrinking_nonlinear}
\end{eqnarray}
where $\mathcal{C}\approx 0.3$ and the dimensionless combination $|p_0| R_0 \kappa_0/k_B T Y_0 \sim (|p_0|/p_c^0) (\ell_\textrm{th}/\ell_\textrm{el})^2$. For sufficiently small outward pressures, the usual linear response term controlled by the bulk modulus $(\mu_0 + \lambda_0)$ is dominated by a nonlinear thermal correction $\sim |p_0|^{\eta/(2-\eta)} \sim |p_0|^{0.67}$. A similar breakdown of Hooke's law appears in the nonlinear response to external tension for thermally fluctuation flat membranes with the same exponent $\eta/(2-\eta)$.~\cite{kosmrlj16} The importance of the nonlinear contribution is determined by the condition $p^* \lesssim 1$, where
\begin{equation}
p^* \equiv \frac{|p_0| R_0 \kappa_0}{k_B T Y_0}.
\label{eq:pstar}
\end{equation}

An alternative renormalization group matching procedure~\cite{rudnick76} also exploits scaling relations such as Eq.~(\ref{eq:scaling_relation}), but instead integrates the recursion relations out to the intermediate scale $\ell^*$  defined by Eq.~(\ref{eq:lstar}), and then matches onto perturbation theory to calculate corrections beyond that scale. We have checked that there are only order of unity differences to the results described here.

\section{Buckling of spherical shells}
\label{sec:buckling_pressure}
\begin{figure}[t]
\includegraphics[scale=.5]{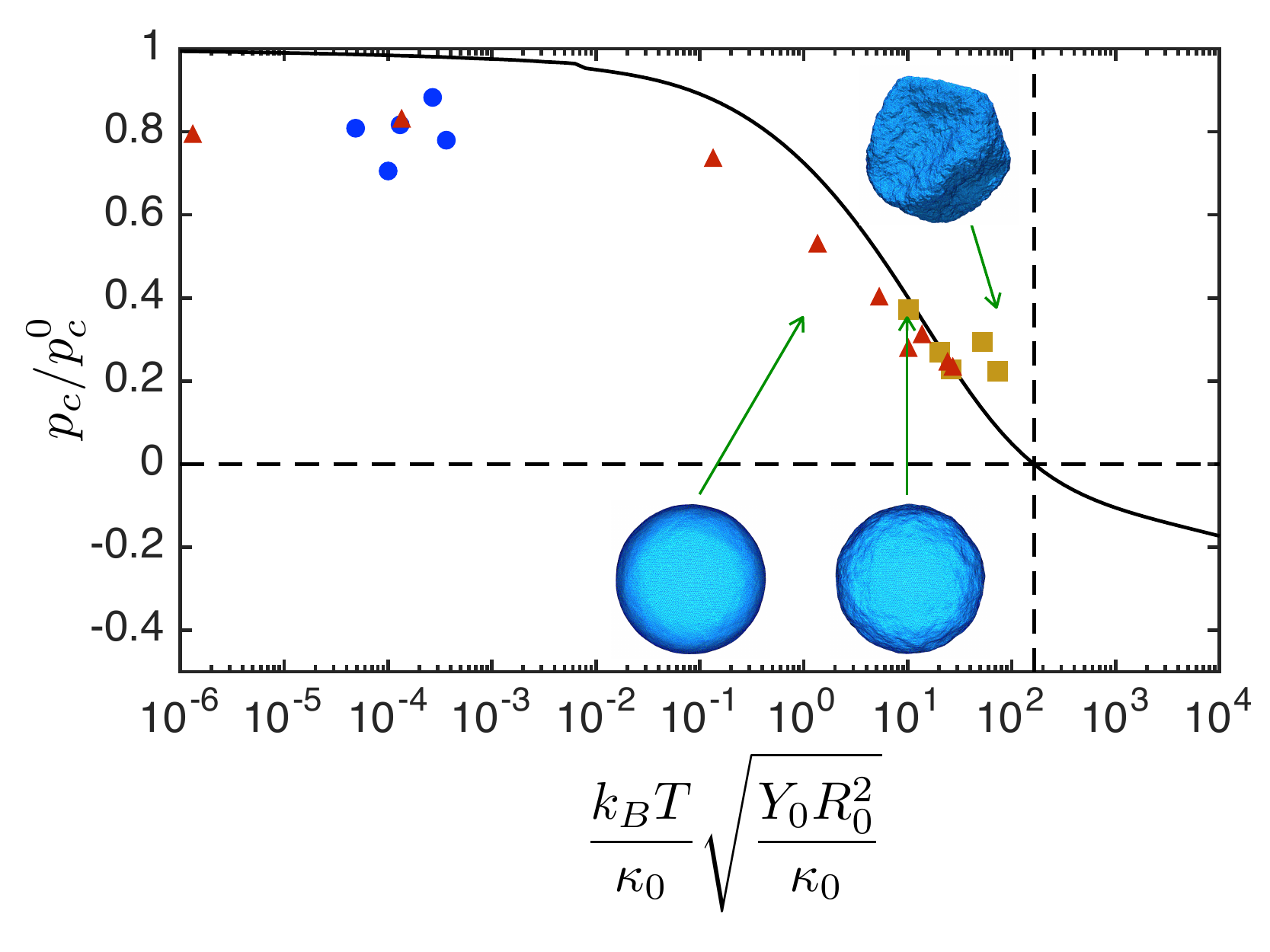}
\caption{(Color online) Thermal fluctuations reduce critical buckling pressure $p_c$ below its classical value $p_c^0$ in Eq.~(\ref{eq:classical_buckling_pressure}), to a point where it can even assume negative values when $(k_B T/\kappa_0) \sqrt{Y_0 R_0^2/\kappa_0} \gg 1$. The solid black line corresponds to the theoretical prediction based on renormalization group calculations and symbols are buckling transitions extracted from the Monte Carlo simulations of Ref.~\cite{paulose12}. Green arrows point to the locations in parameter space $(k_B T/\kappa_0) \sqrt{Y_0 R_0^2/\kappa_0}$ and $p_0/p_c^0$, that correspond to the snapshots of spherical shells from the simulations shown in Fig.~\ref{fig:simulations}. Because for large temperatures $T$ (or equivalently for large shells $R_0$) the critical buckling pressure $p_c$ becomes negative, thermal fluctuations spontaneously crush spherical shells even at zero or somewhat negative external pressures.
}
\label{fig:buckling_pressure}
\end{figure}

By systematically varying the bare external pressure $p_0$ as an initial condition in our renormalization group calculations, we identified the critical buckling pressure $p_c$ for spherical shells in the presence of thermal fluctuations. In agreement with the scaling description embodied in Eqs.~(\ref{eq:scaling_functions}) we found that the critical buckling pressure can be described with a scaling function that depends on a single dimensionless parameter
\begin{equation}
p_c = p_c^0 \ \psi \!\left(\frac{\ell_\textrm{el}}{\ell_\textrm{th}}\right) = p_c^0 \ \Psi \!\left(\frac{k_B T}{\kappa_0} \sqrt{\frac{Y_0 R_0^2}{\kappa_0}} \right),
\label{eq:thermal_buckling_pressure}
\end{equation}
where $\Psi(x)$ is a monotonically decreasing scaling function with
\begin{eqnarray}
\Psi(x) & \approx & \left\{
\begin{array}{c c}
1 -0.28 \, x^{0.4}, & x\ll 1 \\
-0.047 \, x^{(3 \eta-2)/(2+\eta)}, & x \gg 1
\end{array}
\right..
\end{eqnarray}
The small $x$ behavior comes from a fit to our numerical calculations. The $\eta$-dependent power law $\sim - x^{0.14}$ for large  $x$ matches the minimal stabilizing pressure $p_{0,\textrm{min}}$ introduced in Eq.~(\ref{eq:min_p}). Note that thermal fluctuations lead to a substantial reduction in the critical buckling pressure $p_c$ and that $\Psi(x)$ becomes \emph{negative} for $x \gtrsim 160$ (see Fig.~\ref{fig:buckling_pressure}). A remarkable consequence, is that, even when the pressure difference vanishes ($p_0 \equiv 0$), spherical shells are only stable provided they are smaller than
\begin{equation}
R_\textrm{max} \approx 160 \frac{\kappa_0}{k_B T} \sqrt{\frac{\kappa_0}{Y_0}}.
\label{eq:max_radius}
\end{equation}
Larger shells are spontaneously crushed by thermal fluctuations! The condition of zero microscopic pressure difference could be achieved experimentally by studying hemispheres, which should have similar buckling thresholds to spheres, or spheres which (like wiffle balls) have a regular array of large holes. 

The temperature-dependent critical buckling pressures obtained via numerical renormalization group methods are in reasonable agreement with the Monte Carlo simulations of Ref.~\cite{paulose12} (see Fig.~\ref{fig:buckling_pressure}). Note that at small temperatures $T$ and shell sizes $R_0$, where we expect that the critical buckling pressure $p_c$ is approximately equal to the classical buckling pressure $p_c^0$, simulations show systematically lower buckling pressures. This also happens in experiments with macroscopic spherical shells, where the lower buckling pressure is due to shell imperfections~\cite{carlson67}. Similar effects could arise at low temperatures for the amorphous shells simulated in Ref.~\cite{paulose12}. Note that the temperature-dependent critical buckling pressure obtained in this paper were determined by identifying deformation modes, for which the free energy landscape becomes unstable. In practice we expect that even perfectly homogeneous thermalized spherical shells will buckle at a slightly lower external pressure, because the metastable modes embodied in a pressurized sphere exist in a shallow energy minimum, and can escape over a small energy barrier of the order $k_B T$ in the presence of thermal fluctuations.

\section{Conclusions}
\label{sec:conclusions}
In this paper we demonstrated with renormalization group methods that thermal fluctuations in thin spherical shells  become significant when thermal length scale $\ell_\textrm{th}$ [see Eq.~(\ref{eq:thermal_length})] becomes smaller than elastic length scale $\ell_\textrm{el}$ [see Eq.~(\ref{eq:elastic_length})], or equivalently when $(k_B T/\kappa_0) \sqrt{Y_0 R_0^2/\kappa_0} \gtrsim 1$. An identical combination of variables was uncovered in the perturbation calculations of Ref.~\cite{paulose12}. If we assume that shells of thickness $h$ are constructed from a 3D isotropic elastic material with Young's modulus $E_0$ and Poisson's ratio $\nu_0$ [see Eq.~(\ref{eq:elastic_constants_2d_3d})], then the relevant dimensionless parameter can be rewritten as
\begin{equation}
\frac{k_B T}{\kappa_0} \sqrt{\frac{Y_0 R_0^2}{\kappa_0}} = \left[12 (1-\nu_0^2)\right]^{3/2}\frac{k_B T R_0}{E_0 h^4}.
\end{equation}
Thus, this critical dimensionless parameter varies as the inverse $4^\textrm{th}$ power of shell thickness $h$. For thermal fluctuations to become relevant at room temperature, shells only a few nanometers thick may be required. For such shells, thermal fluctuations renormalize elastic constants in the same direction as for flat solid membranes (see Eq.~(\ref{eq:renormalized_elastic_constants}) and Figs.~\ref{fig:RGflows}d-e), i.e. bending rigidity gets enhanced, in-plane elastic constants get reduced and all elastic constants become scale dependent. However, in striking contrast to flat membranes, where an isotropic external tension does not get renormalized,~\cite{kosmrlj16} thermal fluctuations  can strongly enhance the effect of an inward pressure $p_0$. As a consequence, spherical shells  get crushed at a lower external pressure than the classical zero temperature buckling pressure (see Fig.~\ref{fig:buckling_pressure}). In fact, shells that are larger than $R_\textrm{max} \approx 160 (\kappa_0/k_B T) \sqrt{\kappa_0/Y_0}$ become unstable even at zero or slightly negative external pressure. Such large shells can be stabilized by a sufficiently large outward pressure $p_0<0$, which cuts off the renormalization of elastic constants (see Fig.~\ref{fig:RGflows}b). We then find that the shell size increases nonlinearly with internal pressure with a universal exponent characteristic of flat membranes (see Eq.~(\ref{eq:shrinking_nonlinear}) and Fig.~\ref{fig:radius_shrinking}). Note that for sufficiently large outward pressure $p_0 \lesssim  - k_B T Y_0/ R_0 \kappa_0$ the renormalization is completely suppressed and we recover the behavior of classical shells at zero temperature. 

How do these results impact on the physics of currently available microscopic shells? Shells of microscopic organisms come in various sizes and shapes, and they need not be  perfectly spherical. Therefore we just report some characteristic parameters at room temperature $T=300\textrm{K}$, where the radius $R_0$ is identified with half a characteristic shell diameter. For an ``empty'' viral capsid of bacteriophage $\phi29$ (water inside and water outside) with $R_0\approx20\textrm{-}25\textrm{nm}$, $h\approx 1.6 \textrm{nm}$ and $E_0\approx 1.8\textrm{GPa}$,~\cite{ivanovska04} we find that thermal fluctuations have only a small effect [$(k_B T/\kappa_0) \sqrt{Y_0 R_0^2/\kappa_0} \sim 0.3$]. When a capsid of bacteriophage $\phi29$ is filled with viral DNA,  the capsid is under a huge outward osmotic pressure ($p_0 < -6 \textrm{MPa}=-60\textrm{atm}$), which completely suppress thermal fluctuations [$p^* = |p_0| R_0 \kappa_0/(k_B T Y_0) \sim 7$, see Eq.~(\ref{eq:pstar})]. For gram-positive bacteria, which have thick cell wall, thermal fluctuations can be ignored, e.g. for Bacillus subtilis with $R_0\approx 0.4\mu\textrm{m}$, $h\approx 30\textrm{nm}$, $E_0\approx10\textrm{-}50\textrm{MPa}$~\cite{amir14} we obtain $(k_B T/\kappa_0) \sqrt{Y_0 R_0^2/\kappa_0} \sim 10^{-3}$. For gram-negative bacteria with thin cell walls one might think that thermal fluctuations could be important, e.g. for Escherichia coli with $R_0\approx 0.4\mu\textrm{m}$, $h\approx 4\textrm{nm}$, $E_0\approx30\textrm{MPa}$~\cite{amir14} we obtain $(k_B T/\kappa_0) \sqrt{Y_0 R_0^2/\kappa_0}\sim 8$. However, bacteria are under a large outward osmotic stress called turgor pressure, which completely suppresses thermal fluctuations, e.g. for E. coli $p_0 \approx -0.3\textrm{MPa}=-3\textrm{atm}$~\cite{amir14} and dimensionless pressure is $p^*=|p_0| R_0 \kappa_0/(k_B T Y_0) \sim 40 \gg 1$. Note that bacteria regulate osmotic pressure via mechanosensitive channels and hence, they might have evolved to the regime with large turgor pressure in order to protect their cell walls from thermal fluctuations. Somewhat similar to bacteria are nuclei in eukaryotic cells, where genetic material is protected by a nuclear envelope with $R_0 \approx 8 \mu\textrm{m}$, $h/R_0 \sim 10^{-3}\textrm{--}10^{-2}$ and $E_0 \sim 10^{2}\textrm{--}10^4 \textrm{Pa}$,~\cite{kim15} such that $(k_B T/\kappa_0) \sqrt{Y_0 R_0^2/\kappa_0} \sim 10^1\textrm{--}10^7$. When cells are attached to a substrate, densely packed genetic material generates a large outward osmotic pressure $p_0/E_0 \approx - 8\times10^{-2} $, which suppresses thermal fluctuations ($p^*=|p_0| R_0 \kappa_0/(k_B T Y_0) \sim 3\textrm{--}300$). However, upon detachment of cells from the substrate, the cell volume shrinks due to the release of traction forces and the resulting cytoplasm osmotic pressure crushes cell nuclei,~\cite{kim15} a phenomenon that could be influenced by thermal fluctuations.

Thermal fluctuations definitely play an important role in red blood cell membranes. The red blood cell membrane is composed of lipid bilayer with bending rigidity $\kappa_0 \approx 6\textrm{-}40 k_B T$~\cite{park10,evans83} and an attached spectrin network, which contributes to a Young's modulus $Y_0\approx 25 \mu\textrm{N}/\textrm{m}$~\cite{waugh79,park10}, which gives the composite system a resistance to shear. For a characteristic size of $R_0 \approx 7\mu\textrm{m}$ we find $(k_B T/\kappa_0) \sqrt{Y_0 R_0^2/\kappa_0} \approx 2\textrm{-}40$. We neglect here interesting nonequilibrium effects in living cells, where ATP can be burned to turn spectrin into an ``active'' material.~\cite{turlier16} 
Note that by treating red blood cells with mild detergents,  which lyse the cells, one can produce red blood cell ``ghosts'' that are composed of spectrin skeleton alone. Such membranes have smaller bending rigidity and exhibit much larger fluctuations, which was used to confirm the scale-dependence of elastic constants  via  X-ray and light scattering experiments in Ref.~\cite{schmidt93}.

As discussed in Ref.~\cite{paulose12}, artificial microscopic shells have also been constructed from polyelectrolytes~\cite{elsner06}, proteins~\cite{hermanson07} and polymers~\cite{shum08}. Such microcapsules can be made extremely thin, with the thickness of several nanometers, where thermal fluctuations can become relevant. For example, microcapsules with $h\approx 6\textrm{nm}$ thickness were fabricated from reconstituted spider silk~\cite{hermanson07} with $R_0\approx 30 \mu\textrm{m}$ and $E_0\approx 1 \textrm{GPa}$, where we find $(k_B T/\kappa_0) \sqrt{Y_0 R_0^2/\kappa_0} \sim 3$. Similar polymersomes can be made 10 times larger with $R_0\approx 300 \mu{\textrm{m}}$, while being thinner than 10 nanometers.~\cite{shum08} Polycrystalline shells or hemispheres of graphene provide a particularly promising candidate for observing the effects of thermal fluctuations on solid membranes with a spherical background curvature. Indeed, with graphene parameters ($\kappa_0=1.1\textrm{eV}$~\cite{fasolino07} and $Y_0=340\textrm{N}/\textrm{m}$~\cite{lee08}), the maximum allowed radius when $p_0=0$ at room temperature from Eq.~(\ref{eq:max_radius}) is $R_\textrm{max}\approx 160\textrm{nm}$. We hope this paper will stimulate further experimental and numerical investigations of the stability and mechanical properties of thermalized spheres.

\begin{acknowledgments}
We acknowledge support by the National Science Foundation, through grants DMR1306367 and DMR1435999, and through the Harvard Materials Research and Engineering Center through Grant DMR1420570. We would also like to acknowledge useful discussions with Jan Kierfeld and thank Gerrit Vliegenthart for providing snapshots of spherical shells from the Monte Carlo simulations of Ref.~\cite{paulose12}.
\end{acknowledgments}

\appendix
\section{Renormalization group recursion relations for  spherical shells}
\label{sec:app_rg_flow}
In this Appendix we derive the renormalization group recursion relations displayed in Eqs.~(\ref{eq:flow}). We start by rewriting the free energy in Eq.~(\ref{eq:effective_free_energy_both}) in Fourier space as
\begin{subequations}
\label{eq:app:effective_free_energy}
\begin{eqnarray}
F_\textrm{eff} & = & F_0 + F_{\textrm{int}}, \\
\frac{F_0}{A} & = & \sum_{\bf q} \frac{1}{2} \left[ \kappa_0 q^4 - \frac{p_0 R_0 q^2}{2} + \frac{Y_0}{R_0^2}\right] \tilde f({\bf q}) \tilde f(-{\bf q}) \\
\frac{F_\textrm{int}}{A} & = &  \sum_{\substack{{\bf q}_1 + {\bf q}_2 = {\bf q} \ne {\bf 0}\\{\bf q}_3 + {\bf q}_4 = -{\bf q}\ne{\bf 0}}} \frac{Y_0}{8} 
\left[q_{1i} P_{ij}^T({\bf q}) q_{2j} \right] \left[q_{3i} P_{ij}^T({\bf q}) q_{4j} \right] \nonumber\\
&& \quad \quad \quad\quad\quad\quad\quad  \times \tilde f({\bf q}_1) \tilde f({\bf q}_2) \tilde f({\bf q}_3) \tilde f({\bf q}_4) \nonumber \\
 && + \!\!\!\!\sum_{\substack{{\bf q}_1 \ne {\bf 0}\\{\bf q}_2 + {\bf q}_3 = -{\bf q}_1}}
\!\!\!\!  \frac{Y_0}{2 R_0} \left[q_{2i} P_{ij}^T({\bf q}_1) q_{3j} \right]  \tilde f({\bf q}_1) \tilde f({\bf q}_2) \tilde f({\bf q}_3),\nonumber \\
\label{eq:app:effective_free_energy_int}
\end{eqnarray}
\end{subequations}
where $A$ is the area, $\tilde f({\bf q}) = \int (d^2{\bf r}/A) e^{-i {\bf q} \cdot {\bf r}} \tilde f({\bf r})$, and $P_{ij}^T({\bf q}) = \delta_{ij}-q_i q_j/q^2$ is the transverse projection operator. Note that the sums over wavevectors can be converted to integrals in the shallow-shell approximation as $\sum_{\bf q} \rightarrow A \int \!d^2{\bf q}/(2 \pi)^2$.

To implement the momentum shell renormalization group, we first integrate out all Fourier modes in a thin momentum shell $\Lambda/b < q < \Lambda$, where $a=\pi/\Lambda$ is a microscopic cutoff and $b=e^{s}$ with $s \ll 1$. Next we rescale lengths and fields~\cite{aronovitz88, radzihovsky91}
\begin{subequations}
\label{eq:app:rescaling}
\begin{eqnarray}
{\bf r} &=& b {\bf r}', \\
{\bf q} &=& b^{-1} {\bf q}', \\
\tilde f({\bf q}) & = & b^{\zeta_f} \tilde f'({\bf q}'),
\end{eqnarray}
\end{subequations}
where the field rescaling exponent $\zeta_f$ will be chosen to simplify the resulting renormalization group equations. Finally, we define new elastic constants $\kappa'$, $Y'$, and external pressure $p'$, such that the free energy functional in Eq.~(\ref{eq:app:effective_free_energy}) retains the same form after the first two renormalization group steps.

The integration of Fourier modes in a thin momentum shell $\Lambda/b < k < \Lambda$ is formally done with a functional integral
\begin{eqnarray}
F_\textrm{eff}'[\{{\bf q}\}] \!&=& \!-k_B T \ln\!\!\left[\!\int \!\! \mathcal{D}[\tilde f({\bf k})] e^{-(F_0[\{{\bf q},{\bf k}\}] + F_\textrm{int}[\{{\bf q},{\bf k}\}])/k_B T}\right], \nonumber\\
F_\textrm{eff}'[\{{\bf q}\}]& =& F_\textrm{0}[\{{\bf q}\}] -k_B T \ln \left< e^{-F_\textrm{int}[\{{\bf q},{\bf k}\}]/k_B T} \right>_{0,{\bf k}},
\label{eq:app:free_energy_rg}
\end{eqnarray}
where $q<\Lambda/b$ and we introduced the average
\begin{equation}
\left<\mathcal{O}\right>_{0,{\bf k}} = \frac{\int \mathcal{D}[\tilde f({\bf k})] \mathcal{O} e^{-F_0[\{{\bf k}\}]}}{\int \mathcal{D}[\tilde f({\bf k})]e^{-F_0[\{{\bf k}\}]}}.
\end{equation}
The term involving a logarithm in Eq.~(\ref{eq:app:free_energy_rg}) can be expanded in terms of the cumulants
\begin{equation}
F_\textrm{eff}'[\{{\bf q}\}] \!=\!F_\textrm{0}[\{{\bf q}\}] + \sum_{n} \frac{(-1)^{n-1}}{n! (k_B T)^{n-1}} \bigg<\big(F_\textrm{int}[\{{\bf q},{\bf k}\}]\big)^n \bigg>^{(c)}_{0,{\bf k}}, 
\label{eq:app:cumulants}
\end{equation}
where $\langle \mathcal{O}\rangle^{(c)}= \langle \mathcal{O}\rangle$, $\langle \mathcal{O}^2\rangle^{(c)}= \langle \mathcal{O}^2\rangle - \langle \mathcal{O}\rangle^2$, etc. The infinite series in Eq.~(\ref{eq:app:cumulants}) above can be systematically approximated with Feynman diagrams~\cite{amitB}; Fig.~\ref{fig:diagrams} displays all relevant diagrams to one loop order. The contributions of the diagrams in Fig.~\ref{fig:diagrams}c-i are
\begin{widetext}
\begin{subequations}
\begin{eqnarray}
\frac{F_\textrm{eff}'[\{{\bf q}\}]_{(c)}}{A}&=& \sum_{\bf q} \frac{1}{2}\tilde f({\bf q}) \tilde f(-{\bf q}) \int_{\Lambda/b < |{\bf k}| < \Lambda} \frac{d^2{\bf k}}{(2 \pi)^2} A Y G_{ff}\left({\bf k} + \frac{\bf q}{2} \right) \left[q_i P^T_{ij}\left({\bf k} - \frac{\bf q}{2} \right) \big(k_j + \frac{q_j}{2}\big) \right]^2 , \label{eq:app:1}\\
\frac{F_\textrm{eff}'[\{{\bf q}\}]_{(d-g)}}{A} &=& \sum_{\bf q} \frac{1}{2} \tilde f({\bf q}) \tilde f(-{\bf q}) \int_{\Lambda/b < |{\bf k}| < \Lambda} \frac{d^2{\bf k}}{(2 \pi)^2} \frac{(-1) Y^2 A^2}{k_B T R^2} G_{ff}\left({\bf k} + \frac{\bf q}{2} \right) G_{ff}\left({\bf k} - \frac{\bf q}{2} \right) \nonumber\\
&& \quad   \times \left\{ \left[q_i P^T_{ij}\left({\bf k} + \frac{\bf q}{2} \right) \big(k_j - \frac{q_j}{2}\big) \right]^2 -  \left[q_i P^T_{ij}\left({\bf k} - \frac{\bf q}{2} \right) \big(k_j + \frac{q_j}{2}\big) \right] \left[q_i P^T_{ij}\left({\bf k} + \frac{\bf q}{2} \right) \big(k_j - \frac{q_j}{2}\big) \right]
\right.\nonumber \\
&&\quad\quad \left. + 2 \left[q_i P^T_{ij}\left({\bf k} - \frac{\bf q}{2} \right) \big(k_j + \frac{q_j}{2}\big) \right]\left[\big(k_i - \frac{q_i}{2}\big) P^T_{ij}\left({\bf q} \right) \big(k_j + \frac{q_j}{2}\big) \right] + \frac{1}{2} \left[\big(k_i - \frac{q_i}{2}\big) P^T_{ij}\left({\bf q} \right) \big(k_j + \frac{q_j}{2}\big) \right]^2 \right\},\nonumber \\
\label{eq:app:2}\\
\frac{F_\textrm{eff}'[\{{\bf q}\}]_{(h)}}{A} &=&\sum_{\substack{{\bf q} \ne {\bf 0}\\{\bf q}_2 + {\bf q}_3 = - {\bf q}}} \frac{Y}{2R} \left[q_{2i} P_{ij}^T({\bf q}) q_{3j} \right]  \tilde f({\bf q}) \tilde f({\bf q}_2) \tilde f({\bf q}_3)\nonumber \\
&& \quad\quad \times \int_{\Lambda/b < |{\bf k}| < \Lambda} \frac{d^2{\bf k}}{(2 \pi)^2} \frac{(-1) Y A^2}{2 k_B T} G_{ff}\left({\bf k} + \frac{{\bf q}}{2} \right) G_{ff}\left({\bf k} - \frac{{\bf q}}{2} \right) \left[\big(k_i - \frac{q_i}{2}\big) P^T_{ij}\left({\bf q} \right) \big(k_j + \frac{q_j}{2}\big) \right]^2, \label{eq:app:3}\\
\frac{F_\textrm{eff}'[\{{\bf q}\}]_{(i)}}{A} &=& \sum_{\substack{{\bf q}_1 + {\bf q}_2 = {\bf q} \ne {\bf 0}\\{\bf q}_3 + {\bf q}_4 = -{\bf q}\ne{\bf 0}}} \frac{Y}{8} 
\left[q_{1i} P_{ij}^T({\bf q}) q_{2j} \right] \left[q_{3i} P_{ij}^T({\bf q}) q_{4j} \right]  \tilde f({\bf q}_1) \tilde f({\bf q}_2) \tilde f({\bf q}_3) \tilde f({\bf q}_4) \nonumber \\
&& \quad\quad \times \int_{\Lambda/b < |{\bf k}| < \Lambda} \frac{d^2{\bf k}}{(2 \pi)^2} \frac{(-1) Y A^2}{2 k_B T} G_{ff}\left({\bf k} + \frac{{\bf q}}{2} \right) G_{ff}\left({\bf k} - \frac{{\bf q}}{2} \right) \left[\big(k_i - \frac{q_i}{2}\big) P^T_{ij}\left({\bf q} \right) \big(k_j + \frac{q_j}{2}\big) \right]^2, \label{eq:app:4}
\end{eqnarray}
\end{subequations}
where $G_{ff}({\bf q}) = k_B T/[A(\kappa q^4 - p R q^2/2 + Y/R^2)]$, and subscripts $(c)$, $(d-g)$, $(h)$ and $(i)$ describe contributions from the corresponding diagrams in Fig.~\ref{fig:diagrams}. The integrands in the equations above must now be expanded for small wavevectors ${\bf q}$. The relevant contributions to $\kappa'$, $p'$ and $Y'$ are related to terms that scale with $q^4$, $q^2$ and $q^0$ in Eqs.~(\ref{eq:app:1}) and (\ref{eq:app:2}), respectively. The contributions to three-point and four-point vertices are described with Eqs.~(\ref{eq:app:3}) and (\ref{eq:app:4}), respectively, and here it is enough to keep only the $q^0$ terms in the integrands.

After the integration of Fourier modes in a thin momentum shell $\Lambda/b < k < \Lambda$, where $b=e^s$ with $s\gg 1$, rescaling fields, momenta and lengths according to Eq.~(\ref{eq:app:rescaling}) we find the recursion relations
\begin{subequations}
\label{eq:app:recursion_relations}
 \begin{eqnarray}
 \beta_\kappa  &=&  \frac{d \kappa'}{d s} = 2 (\zeta_f - 1) \kappa' + \frac{3 k_B T Y' \Lambda^2}{16 \pi \mathcal{D}} - \frac{3 k_B T Y'^2 \Lambda^2}{8 \pi R'^2 \mathcal{D}^2} \bigg[1 +\frac{I_{\kappa1}}{\mathcal{D}^2} + \frac{I_{\kappa2}}{\mathcal{D}^4} \bigg],\\
 \beta_Y &=&  \frac{d Y'}{d s} = 2 \zeta_f Y' - \frac{3 k_B T Y'^2 \Lambda^6}{32 \pi  \mathcal{D}^2},  \label{eq:app:renormalization_flow_Y}\\
 \beta_{p}  &=&  \frac{d p'}{d s}= (2 \zeta_f +1) p' + \frac{3 k_B T Y'^2 \Lambda^4}{4 \pi R'^3 \mathcal{D}^2} \left[1 + \frac{I_p}{\mathcal{D}^2} \right], \quad \  \\
 \beta_{R}  &=&  \frac{d R'}{d s} = - R',
 \end{eqnarray} 
 \end{subequations}
 where we introduce a denominator factor $\mathcal D$ and the results of various integrations as
 \begin{subequations}
 \label{eq:app:rg_flow_detailed}
  \begin{eqnarray}
  \mathcal{D}&=& \kappa' \Lambda^4 - \frac{p' R' \Lambda^2}{2} + \frac{Y'}{R'^2}, \\
  I_{\kappa1} & = & \frac{1}{48} \bigg[ - \frac{4 Y'^2}{R'^4} + 8 \frac{Y'}{R'^2} \left(2 p' R' \Lambda^2 - 9 \kappa' \Lambda^4\right) - \left(5 p'^2 R'^2 \Lambda^4 - 32 p' R' \kappa \Lambda^6 + 36 \kappa'^2 \Lambda^8 \right)\bigg], \\
  I_{\kappa2} & = & \frac{1}{768} \bigg[ -\frac{24 Y'^3 \kappa' \Lambda^4}{R'^6}  + \frac{Y'^2}{R'^4} \big(9 p'^2 R'^2 \Lambda^4- 76 p' R' \kappa' \Lambda^6 +  268 \kappa'^2 \Lambda^8 \big) \nonumber \\
  && \quad \quad + \frac{Y'}{R'^2} \big(-5 p'^3 R'^3 \Lambda^6 + 52 p'^2 R'^2 \kappa' \Lambda^8   -204 p' R' \kappa'^2 \Lambda^{10} + 
 160 \kappa'^3 \Lambda^{12} \big) \nonumber \\
 && \quad \quad + \big( p'^4 R'^4 \Lambda^8 - 12 p'^3 R'^3 \kappa' \Lambda^{10} + 
 56 p'^2 R'^2 \kappa'^2 \Lambda^{12} - 96 p' R' \kappa'^3 \Lambda^{14} +  60 \kappa'^4 \Lambda^{16} \big) \bigg],  \\
 I_p & = & \frac{1}{48} \bigg[ \frac{Y}{R^2} \big(3 p' R' \Lambda^2 - 16 \kappa' \Lambda^4\big) + \big( -p'^2 R'^2 \Lambda^4 + 7 p' R' \kappa' \Lambda^6 - 
 8 \kappa'^2 \Lambda^8\big) \bigg].
   \end{eqnarray} 
  \end{subequations}
  \end{widetext}
The $\beta_Y$ recursion relation in Eq.~(\ref{eq:app:renormalization_flow_Y}) describes changes in the quadratic ``mass'' proportional to $Y$ in Eq.~(\ref{eq:app:effective_free_energy}). Similarly, we can calculate the recursion relations for the cubic and quartic terms in Eq.~(\ref{eq:app:effective_free_energy}). The only significant change is in the effect of rescaling: the $2 \zeta_f Y$ term now becomes $(3 \zeta_f - 1) Y$ and $(4 \zeta_f -2) Y$, respectively.

  \section{Independence of renormalization group results on the choice of $\zeta_f $}
  \label{sec:app:rg_flow_sheets}
  In this section we illustrate the insensitivity of the renormalization procedure to the precise choice of the field rescaling factor that appears in $\tilde f({\bf q})=b^{\zeta_f} \tilde f'({\bf q}')$. Specifically we demonstrate that for a flat thermalized sheet we show that the renormalized bending rigidity $\kappa_R(\ell)$ and renormalized Young's modulus $Y_R(\ell)$ are identical, when we chose either $\zeta_f(s)\equiv1$, as we did for convenience with spherical shells, or we choose $\zeta_f(s)$ such that the the $\kappa'(\ell)\equiv\kappa_0$ remains fixed, as is the case in the usual renormalization group procedure.~\cite{radzihovsky91}
  
  The recursion relations for flat sheets are~\cite{radzihovsky91,kosmrlj16} 
  \begin{subequations}
 \label{eq:appB:flow}
 \begin{eqnarray}
 \beta_\kappa  &=&  \frac{d \kappa'}{d s} = 2 (\zeta_f - 1) \kappa' + \frac{3 k_B T Y'}{16 \pi \kappa' \Lambda^2},  \label{eq:appB:flow_kappa}\\
 \beta_Y &=&  \frac{d Y'}{d s} = (4 \zeta_f - 2) Y' - \frac{3 k_B T Y'^2}{32 \pi  \kappa'^2 \Lambda^2}.
 \end{eqnarray} 
 \end{subequations}
The scale-dependent parameters $\kappa'(s)$, $Y'(s)$, which are obtained by integrating the differential equations in Eqs.~(\ref{eq:appB:flow}) up to $s=\ln(\ell/a)$ with initial conditions $\kappa'(0)=\kappa_0$, $Y'(0)=Y_0$, are related to the scaling of propagator $G_{ff}(q)$ according to~\cite{amitB}
\begin{equation}
G_{ff}(q|\kappa_0, A) =e^{\int \!2 \zeta_f(s) ds} G_{ff}(q e^s | \kappa'(s), A e^{-2 s}),
\label{eq:app:scaling_relation}
\end{equation}
where $G_{ff}(q|\kappa,A)=k_B T/[A \kappa q^4]$ and we explicitly wrote the rescaled momenta $q'=q e^s$ and the rescaled patch area $A'=Ae^{-2 s}$. By replacing the left hand side in the Eq.~(\ref{eq:app:scaling_relation})  above with the propagator $G_{ff}(q)=k_B T/[A (\kappa_R(q) q^4)]$, we find the renormalized bending rigidity
\begin{eqnarray}
\kappa_R (s) & = & \kappa' (s) e^{\int[2-2 \zeta_f(s)] ds}.
\label{eq:app:kappaR}
\end{eqnarray}
From a similar scaling relation for the four-point vertex we find 
\begin{eqnarray}
Y_R (s) & = & Y' (s) e^{\int [2 - 4\zeta_f(s)] ds}.
\label{eq:app:YR}
\end{eqnarray}

\begin{figure*}[t]
\includegraphics[scale=.5]{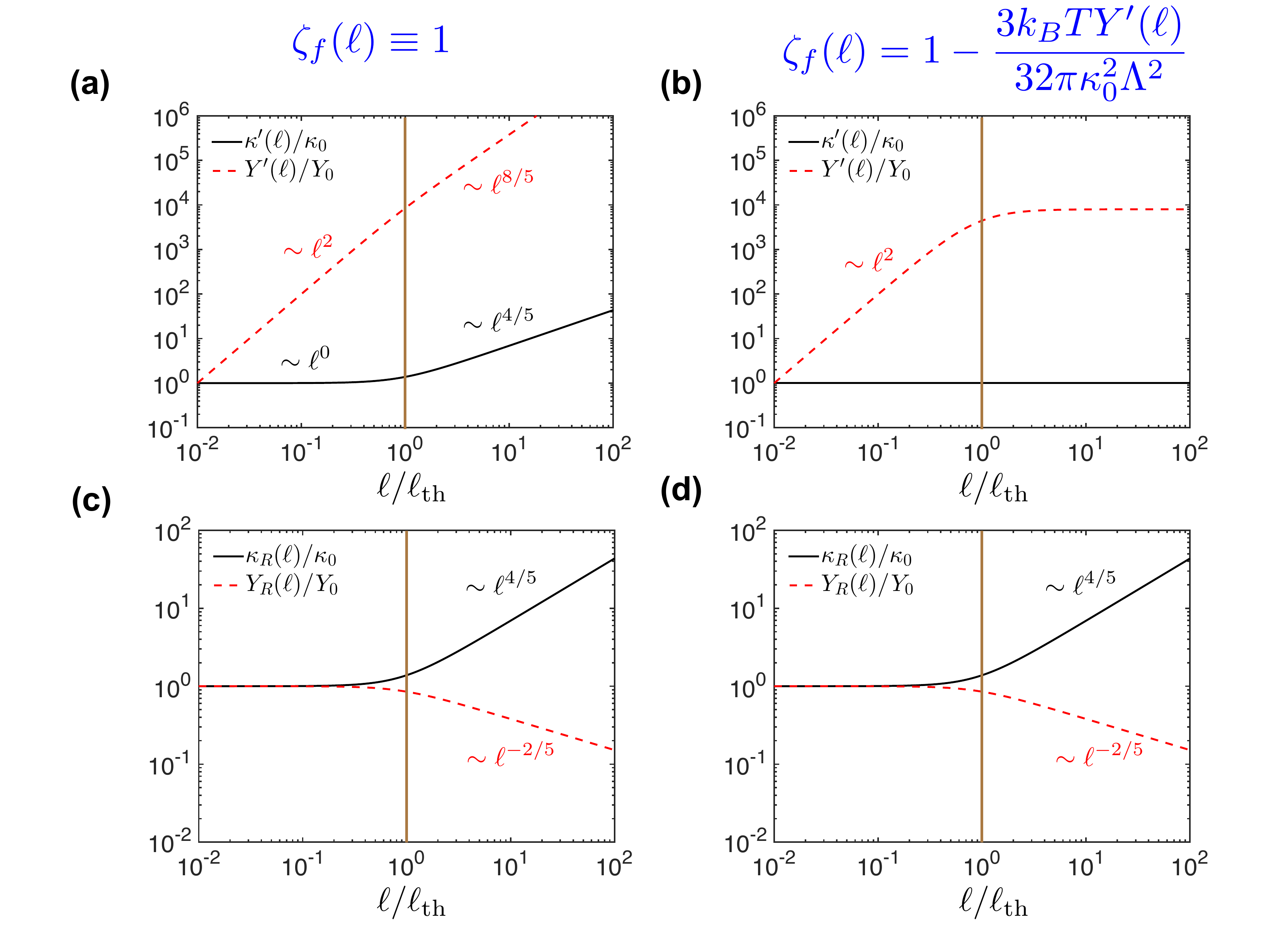}
\caption{(Color online) Renormalization group flows in thermalized flat sheets with $a/\ell_\textrm{th}=10^{-2}$ for two different choices of scaling exponents $\zeta_f$. Plots on the left correspond to $\zeta_f(\ell)\equiv 1$ and plots on the right correspond to $\zeta_f(\ell)=1-3k_B T Y'(\ell)/(32 \pi \kappa_0^2 \Lambda^2)$, which fixes $\kappa'(\ell) \equiv \kappa_0$. (a-b) Renormalization group flows for $\kappa'(\ell)$ and $Y'(\ell)$  obtained (a) from Eq.~(\ref{eq:appB:flow}) and (b) from Eq.~(\ref{eq:appB:flow2}). (c-d) Scale dependence of renormalized elastic constants $\kappa_R(\ell)$ and $Y_R(\ell)$ obtained by removing the scaling factors from $\kappa'(\ell)$ and $Y'(\ell)$ as described in Eqs.~(\ref{eq:app:kappaR}) and (\ref{eq:app:YR}). Note that the physical renormalized constants $\kappa_R(\ell)$ and $Y_R(\ell)$ are identical in (c) and (d), even though the flows of $\kappa'(\ell)$ and $Y'(\ell)$ in (a) and (b) depend on the precise choice of the scaling exponent $\zeta_f(\ell)$.
}
\label{fig:app}
\end{figure*}

First we choose $\zeta_f(s)\equiv 1$, which leads to the recursion relations to 
  \begin{subequations}
 \begin{eqnarray}
  \frac{d \kappa'}{d s}& =& \frac{3 k_B T Y'}{16 \pi \kappa' \Lambda^2},\\
 \frac{d Y'}{d s} &= & 2 Y' - \frac{3 k_B T Y'^2}{32 \pi  \kappa'^2 \Lambda^2}.
 \end{eqnarray} 
 \label{eq:appB:flow}
 \end{subequations}
 By integrating the differential equations in Eqs.~(\ref{eq:appB:flow}) up to $s=\ln(\ell/a)$ with initial conditions $\kappa'(0)=\kappa_0$ and $Y'(0)=Y_0$ we find (see Fig.~\ref{fig:app})
 \begin{subequations}
 \begin{eqnarray}
 \kappa'(\ell)& \sim& \left\{
 \begin{array}{c c}
 \kappa_0, & \ell \ll \ell_\textrm{th} \\
 \ \  \quad \kappa_0 (\ell/\ell_\textrm{th})^{4/5},\quad \ \ \,  & \ell \gg \ell_\textrm{th}
 \end{array}
 \right. ,\\
Y'(\ell) &\sim& \left\{
 \begin{array}{c c}
 Y_0 (\ell/a)^2, & \ell \ll \ell_\textrm{th} \\
 Y_0 (\ell_\textrm{th}/a)^2 (\ell/\ell_\textrm{th})^{8/5}, & \ell \gg \ell_\textrm{th}
 \end{array}
 \right.,
\end{eqnarray}
\end{subequations}
where $\ell_\textrm{th}\sim \kappa_0/\sqrt{k_B T Y_0}$. Upon removing scaling factors according to Eqs.~(\ref{eq:app:kappaR}) and (\ref{eq:app:YR}) we obtain our final scale-dependent renormalized elastic constants
 \begin{subequations}
\label{eq:appB:renormalized_constants}
 \begin{eqnarray}
 \kappa_R(\ell)& \sim& \left\{
 \begin{array}{c c}
 \kappa_0, & \ell \ll \ell_\textrm{th} \\
\ \kappa_0 (\ell/\ell_\textrm{th})^{4/5}, \ & \ell \gg \ell_\textrm{th}
 \end{array}
 \right. ,\\
Y_R(\ell) &\sim& \left\{
 \begin{array}{c c}
 Y_0, & \ell \ll \ell_\textrm{th} \\
 Y_0(\ell/\ell_\textrm{th})^{-2/5}, & \ell \gg \ell_\textrm{th}
 \end{array}
 \right.,
\end{eqnarray}
\end{subequations}
where we recognize the usual scaling exponents $\eta=4/5$ and $\eta_u=2/5$, which satisfy identity $\eta_u + 2 \eta=2$.

A more conventional choice,~\cite{radzihovsky91, aronovitz88} is to take $\zeta_f(s)$ such that the $\kappa'(s)\equiv \kappa_0$ remains fixed. Upon setting $\beta_\kappa=0$ in Eq.~(\ref{eq:appB:flow_kappa}) we find
  \begin{subequations}
\begin{eqnarray}
 \label{eq:appB:flow2}
\zeta_f(s) &=& 1 - \frac{3 k_B T Y'(s)}{32 \pi \kappa_0^2 \Lambda^2} \\
\frac{d Y'(s)}{d s} &=& 2 Y'(s) - \frac{15 k_B T Y'(s)^2}{32 \pi  \kappa_0^2 \Lambda^2}, 
\end{eqnarray}
  \end{subequations}
 By integrating the differential equations in Eqs.~(\ref{eq:appB:flow2}) up to $s=\ln(\ell/a)$ with initial condition $Y'(0)=Y_0$ we find a fixed point, which is reached at the thermal scale, $\ell \sim \ell_\textrm{th}$ (see Fig.~\ref{fig:app}) such that
  \begin{subequations}
 \begin{eqnarray}
\zeta_f(\ell)& \sim& \left\{
 \begin{array}{c c}
1, & \ell \ll \ell_\textrm{th} \\
\quad \quad  \frac{3}{5},  \quad \quad & \ell \gg \ell_\textrm{th}
 \end{array}
 \right. ,\\
Y'(\ell) &\sim& \left\{
 \begin{array}{c c}
 Y_0 (\ell/a)^2, & \ell \ll \ell_\textrm{th} \\
\  \ \frac{64 \pi \kappa_0^2 \Lambda^2}{15 k_B T}, \ \ & \ell \gg \ell_\textrm{th}
 \end{array}
 \right..
\end{eqnarray}
\end{subequations}
By taking into account scaling factors in Eqs.~(\ref{eq:app:kappaR}) and (\ref{eq:app:YR}), it is easy to see that the value of exponent $\zeta_f^* = 3/5$ at the fixed point leads to the scaling exponents $\eta=2-2 \zeta_f^*=4/5$ and $\eta_u = 4 \zeta_f^* - 2 = 2/5$. From these relations one also finds the identity $\eta_u + 2\eta =2$ regardless of the precise value of $\zeta_f^*$. From Fig.~(\ref{fig:app})  we see that the renormalized bending rigidity $\kappa_R(\ell)$ and the renormalized Young's modulus $Y_R(\ell)$ are identical to the ones obtained in Eq.~(\ref{eq:appB:renormalized_constants}) with the choice of $\zeta(s)\equiv 1$.

\bibliography{library}
\end{document}